\documentclass[12pt,a4paper]{article}
\pdfoutput=1
\usepackage{jheppub}
\usepackage{epstopdf}
\usepackage{graphicx}
\usepackage{epsfig}
\usepackage{dcolumn}  
\usepackage{bm}    
\usepackage{amssymb} 
\usepackage{amsmath,bm}
\usepackage{amsfonts}    
\usepackage{slashed}  
\usepackage[mathscr]{euscript}
\usepackage{epsfig}
\usepackage[position=t,singlelinecheck=off]{subfig}
\usepackage{caption}
\begin{document}
\def\be{\begin{equation}}
\def\ee{\end{equation}}
\def\bea{\begin{eqnarray}} 
\def\eea{\end{eqnarray}}

\title{Nonequilibrium dynamics of the $O(N)$ model on dS$_3$ and AdS crunches}
\author{S. Prem Kumar$^{a}$}  
\author{and  Vladislav Vaganov$^{b}$}
\affiliation{$^{a}$Department of Physics,\\ Swansea University, \\
   Singleton Park, Swansea \\SA2 8PP, UK.\\}

\affiliation{$^{b}$Fields, Gravity \& Strings, \\Center for Theoretical Physics of the Universe,\\
   Institute for Basic Science (IBS), \\ 55 Expo-ro, Yuseong-gu, Daejeon 34126, Korea.
   }

\emailAdd{s.p.kumar@swansea.ac.uk, vlad.vaganov@gmail.com}
\date{\today}
\abstract{ We study the nonperturbative quantum evolution of the interacting $O(N)$  vector model at large-$N$, formulated on a spatial two-sphere, with   {\em time dependent} couplings which diverge at finite time. This model - the so-called ``E-frame" theory, is related via a conformal transformation to the interacting $O(N)$ model in three dimensional global de Sitter spacetime with time {\em independent} couplings. We show that with a purely quartic, relevant deformation the quantum evolution of the E-frame model is regular even when the classical theory is rendered singular at the end of time by the diverging coupling. Time evolution drives the E-frame theory to the large-$N$ Wilson-Fisher fixed point when the classical coupling diverges. We study the quantum evolution numerically for a variety of initial conditions and demonstrate the finiteness of the energy at the classical ``end of time".  With an additional (time dependent) mass deformation, quantum backreaction  lowers the mass, with a putative smooth time evolution only possible in the limit of infinite quartic coupling. We discuss the relevance of these results for the resolution of crunch singularities in AdS geometries dual to E-frame theories with a classical gravity dual. 
}
\maketitle
\section{Introduction}
Revealing possible mechanisms for resolving (spacelike) cosmological singularities in classical gravity is one of the major goals of any consistent microscopic framework  for  gravity. This continues to be a challenge for descriptions of gravity involving the AdS/CFT correspondence and holography where quantum gravity in asymptotically AdS spacetimes is dual to a large-$N$ field theory on the conformal boundary of such a spacetime \cite{maldacena, Witten:1998qj, gkp}. 

Intriguingly, there exist several examples of asymptotically AdS backgrounds exhibiting an FRW crunch singularity, dual to deformations of large-$N$ CFTs placed in de Sitter spacetime \cite{hh, turokcraps1, turokcraps2, maldacena2, br1, br2, paper1, paper2}. Such ``crunching-AdS" geometries arise in de-Sitter-sliced, asymptotically AdS spacetimes where the crunch singularity is cloaked behind a bulk horizon. Despite  superficial resemblance to the crunches in the interior of AdS black holes, the crunching geometries in question are of a qualitatively distinct nature. They occur in globally  defined asymptotically AdS geometries, wherein Hamiltonian evolution of the boundary QFT ends at a singularity in finite time. At this ``end of time", the bulk crunch singularity meets and engulfs the boundary. As emphasized in the works \cite{br1} and \cite{br2}, this can be seen via a simple ``complementarity map" in the bulk which reduces to a conformal transformation on the boundary. The transformation in question maps a CFT plus deformations on a background de Sitter spacetime in $d$ dimensions to the same CFT on $S^{d-1}\times {\mathbb R}_\tau$, but with {\em time dependent} deformation parameters:
\be
{\cal L}_{\rm CFT}\,+\,\left. \sum_i\lambda_{i} \,{\cal O}_i\quad\right|_{{\rm dS}_d}\,\to\,
 {\cal L}_{\rm CFT}\,+\,\sum_i\left.\frac{\lambda_{i}}{\left(\cos\tau\right)^{d-\Delta_i}} \,{\cal O}_i\quad\right|_{S^{d-1}\times {\mathbb R}_\tau}\label{map}
\ee  
where $\Delta_i$ are the conformal dimensions of operators ${\cal O}_i$ and the global de Sitter time $t$ related to conformal time as $\cos\tau\,=\,\left(\cosh\frac{t}{R}\right)^{-1}$.
\begin{figure}[h]
\centering
\includegraphics[width=2.5in]{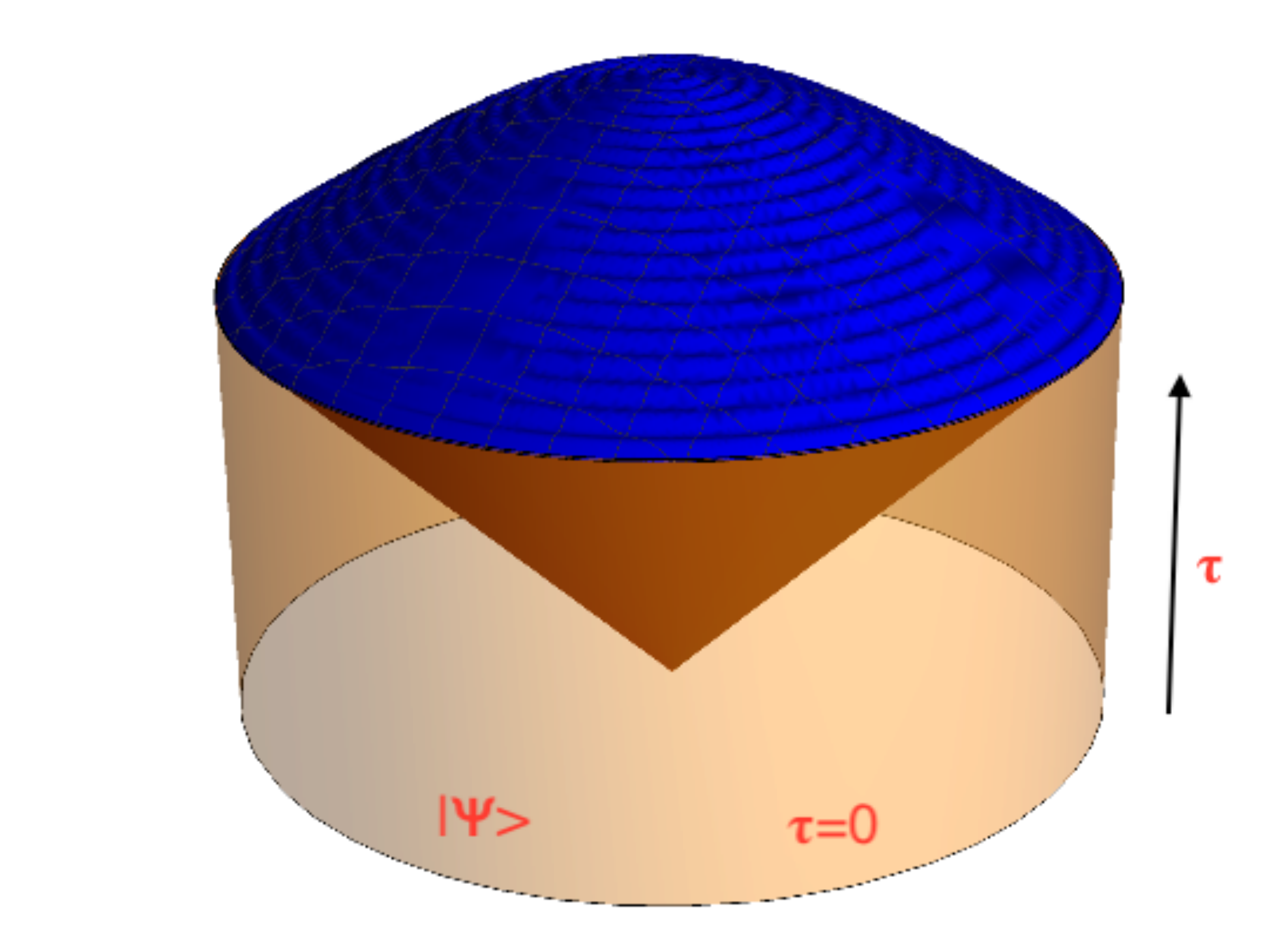}\caption{\small{A depiction of an asymptotically globally AdS geometry, with (E-frame) time evolution ending in a crunch (blue), dual to a boundary CFT in global de Sitter space with relevant deformations. The cone represents the horizon which only appears in the dS-sliced geometry.}}
\label{dscrunch}
\end{figure}

For relevant deformations, the map \eqref{map} renders the classical theory singular at the end of time $\tau=\frac{\pi}{2}$, when the corresponding couplings diverge. In cases where an appropriate large-$N$ limit exists, the theory on the static Einstein space $S^{d-1}\times {\mathbb R}_\tau$, henceforth called the ``E-frame" theory \cite{br1}, is dual to gravity in asymptotically global AdS$_{d+1}$ spacetime.  The appearance of a bulk crunch singularity may then be viewed as a singularity in the time evolution of the boundary field theory driven by  the divergent classical couplings.
Note that while the de Sitter evolution is perfectly smooth and well defined for all times, evolution by the E-frame Hamiltonian is generically expected to be singular.

An important question then is whether, and under what circumstances, is the bulk crunch singularity potentially resolvable within a consistent microscopic framework of the bulk gravity theory. Various approaches towards answering this and similar questions have been proposed and investigated  \cite{turokcraps1, turokcraps2,  br1,  br2, Smolkin:2012er, Sezgin:2005hf, Das:2006dz, Awad:2008jf, Auzzi:2012ca, Engelhardt:2015gla} from both QFT and holographic perpsectives. Holographic probes of such geometries and the associated  bulk singularities have also been considered in order to arrive at a precise interpretation of the singularities \cite{Engelhardt:2013tra, paper1, paper2, Bzowski:2015clm}, inspired by analogous investigations of the black hole singularity in AdS/CFT \cite{Fidkowski:2003nf, Festuccia:2005pi}.

In this paper, we study the setup described above from a purely field theoretic standpoint. In particular, we are interested in understanding whether a CFT with driven
relevant deformations of the kind shown in eq.\eqref{map} can have a smooth evolution towards the end of time when the couplings diverge.  We examine the interacting $O(N)$ vector model in three dimensions in the large-$N$ limit,  which is an exactly solvable theory, and also posesses a classical, higher spin gravity dual \cite{kp, vasiliev}. The free theory has two relevant $O(N)$-singlet deformations: a mass term which has conformal dimension $\Delta=1$ and the quartic interaction with $\Delta =2$. We consider  both these deformations with the field theory on three dimensional de Sitter spacetime (dS$_3$), related by the conformal map \eqref{map} to the classically singular E-frame 
description.

The large-$N$ limit of the interacting $O(N)$ model is controlled by a saddle point \cite{Moshe:2003xn, ZinnJustin:2002ru} which resums superdaisy/cactus diagrams. The saddle point conditions are nonlinear, self-consistent equations which control the time evolution of the large-$N$ field theory treated as an initial value problem. We prepare the field theory in a time-independent, equilibrium state on $S^2$  for times $\tau <0$ and then subject the system to a ``quench" by switching on  time dependent couplings for $\tau \geq 0$ as dictated by the E-frame Lagrangian \eqref{map}.  This approach to large-$N$ dynamics following a quench has been extensively used in a variety of related physical situations (see e.g. \cite{Cooper:1994hr, Boyanovsky:1994me, Boyanovsky:1996rw, Boyanovsky:1997xt}). Our main conclusions are:
\begin{itemize}
\item{The theory with a purely quartic interaction ($\Delta=2$ in $d=3$ and thus relevant), where the renormalized mass of the theory is tuned to vanish, is non-conformal at $\tau=0$, but subsequently driven to the large-$N$ Wilson-Fisher critical point at $\tau=\frac{\pi}{2}$.  Since the theory is conformal at this point, it can be smoothly evolved through, even though the classical Lagrangian is singular. We demonstrate this first for the theory prepared in the vacuum state at $\tau=0$ and then more nontrivially, by numerical evolution, in cases where the initial state is a finite energy excited configuration. In the latter situation we show that E-frame energy per degree of freedom remains finite throughout the quantum evolution, including at $\tau=\frac{\pi}{2}$. Specifically, the system is gapped (on $S^2$) and the gap remains finite in E-frame at all times. This is in stark contrast to the energy of the classical system which necessarily diverges at  $\tau=\frac{\pi}{2}$.}
\item{The inclusion of a mass deformation in addition to the quartic deformation has a significant effect. We find in this case that the large-$N$ time evolution generically drives the quantum fluctuations in the theory to reduce, but not completely cancel, the magnitude of the effective mass in the de Sitter picture. This means that the E-frame theory continues to be singular at $\tau=\frac{\pi}{2}$ because the energy per degree of freedom actually diverges at this time as $\sim \left(\frac{\pi}{2}-\tau\right)^{-2}$, (equivalently, the gap diverges with time) as would be expected from the classical Lagrangian \eqref{map} with a mass deformation ($\Delta=1$ in $d=3$). In the limit of infinite quartic coupling however, we obtain numerical evidence that the effective mass is driven to zero by time evolution in the quantum theory in de Sitter frame, suggesting a smooth time evolution of the corresponding E-frame theory.}
\item{In both types of situations described above, we also consider initial conditions and parameters which render the initial effective mass-squared of excitations sufficiently {\em negative} so as to trigger instabilities. The nonperturbative resummation of quantum effects incorporated within the large-$N$ limit ensures that the growth of quantum fluctuations backreacts on and stabilizes the system at late times, curing any putative instabilities. In addition, the finite volume of the $S^2$ precludes symmetry breaking and thus the theory is always realized in an $O(N)$-symmetric stable phase.}
\end{itemize}
One immediate inference following from these observations would be that relevant deformations which drive a given CFT to a new fixed point should lead to smooth evolution of the quantum theory in the E-frame description despite a diverging, time dependent microscopic coupling.  In particular, the dual gravitational description of such a situation should be devoid of crunch singularities. We hasten to add that the $O(N)$ model also has the important feature that both its critical points (the free theory and the large-$N$ version of the Wilson-Fisher point) are endowed with an infinite set of higher spin conserved currents. Therefore, both ingredients i.e. an approach to a conformal point and appearance of higher spin currents may be necessary to avoid bulk crunch singularities within a dual gravitational setting. This is of particular relevance to  FRW crunch singularities occurring within supergravity duals of large-$N$ CFTs placed in de Sitter spacetime with relevant operators turned on (and with a stable ground state for the  resulting theory) \cite{maldacena2, paper1, paper2}. Within Einstein gravity, the theorem of Abbott and Coleman \cite{Abbott:1985kr} implies that any deformation of (dS-sliced) AdS will lead to a crunch singularity. This would also include any putative deformation that drives the original CFT towards an IR fixed point.  In such a situation we would expect the bulk crunch to be resolved by the inclusion of higher spin or stringy corrections.

The paper is organized as follows: We begin by laying out our conventions and notation to set up the $O(N)$ model in de Sitter spacetime, and its transformation to $S^2 \times {\mathbb R}_\tau$ or ``E-frame", in sections 2 and 3.  In section 4 we review the large-$N$ limit and write down the ensuing saddle-point conditions as evolution equations in real time, suitable for addressing initial value problems. This also includes a discussion of renormalization and cutoff dependence which are relevant for numerical evolution of the system. Section 5 is devoted to the detailed study of the massless theory with the quartic, relevant deformation. Both analytic and numerical results are used to establish the smoothness of the time evolution of this system in E-frame.  In section 6, we turn to a detailed analysis of the massive, interacting theory.  In section 7 we verify that the evolution of the system (in the de Sitter picture),  with the initial conditions we employ, is consistent with an approach to the standard Bunch-Davies vacuum at late times.
\section{$O(N)$ model on dS$_3$}

The $O(N)$  vector model at large-$N$ is exactly solvable. On ${\mathbb R}^3$, the theory exhibits two critical points: the free fixed point in the UV and an IR fixed point which is the large-$N$ version of the Wilson-Fisher fixed point.  We are interested in the dynamics of this theory in de Sitter spacetime in the spherical slicing or global coordinate system:
\be
ds^2\,=\,-dt^2\,+\,R^2\cosh^2\left(\tfrac{t}{R}\right)\,d\Omega_2^2\,.\label{dsglobal}
\ee
In conformal time, defined as,
\be
dt\,=\, R\cosh\left(\tfrac{t}{R}\right)\,d\tau\,\implies\,\cos\tau\,=\, \frac{1}{\cosh\left(\tfrac{t}{R}\right)}\,,
\ee
the metric is conformal  to the Einstein static universe (ESU) or $S_2\times  {\mathbb R}_\tau$:
\be
ds^2\,=\,\frac{R^2}{\cos^2\tau} \left(-d\tau^2\,+\,d\Omega_2^2\right)\,,\qquad
-\frac{\pi}{2} < \tau < \frac{\pi}{2}\,.\label{esumetric}
\ee
Our goal will be to understand the fate of real time dynamics of the $O(N)$ model in the conformal time or ESU picture which will be called the  ``E-frame".  The de Sitter metric in E-frame \eqref{esumetric}, taken at face value, exhibits a divergent conformal factor at $\tau=\pm \frac{\pi}{2}$.  We want to determine whether, and under what circumstances if any, the time evolution of the {\em interacting} $O(N)$ model treated as an initial value problem at $\tau=0$, can be continued meaningfully to (and possibly, past) the ``end of time" at $\tau = \frac{\pi}{2}$. 


Our starting point is the interacting $O(N)$ model defined on ${\rm dS}_3$ which will be called the dS-frame picture. We consider the $N$-tuplet of scalars 
\be
\vec \Phi\,=\,\left(\phi_1, \phi_2\ldots \phi_N\right),
\ee
conformally coupled to the scalar curvature of a fixed background de Sitter spacetime in $2+1$ dimensions in the global coordinate system \eqref{dsglobal}.
The $O(N)$-invariant Lagrangian density for $N$ conformally coupled scalars on this background with quartic interactions is given by:
\bea
&&{\cal L}_{\rm dS_3}\,=\,R^2 \cosh^2 \left(\tfrac{t}{R}\right)\times\\\nonumber
&&\left[\tfrac{1}{2}\left(\partial_t\vec\Phi\right)^2\,-\,\tfrac{1}{2}\,R^{-2}\,{\rm sech}^2 \left(\tfrac{t}{R}\right)\,\left({\bf \nabla}_{{\rm S}^2}\,\vec \Phi\right)^2\,-\,\tfrac{1}{2}\left(M^2_0\,+\,\xi\,{\cal R}\right)\vec \Phi\cdot\vec\Phi\,-\,\frac{\lambda}{4 N}\left(\vec \Phi\cdot\vec \Phi\right)^2\right]\,.
\eea
The conformal coupling $\xi\equiv (d-2)/4(d-1)$ in $d$ spacetime dimensions to the background de Sitter scalar curvature ${\cal R}$ results in a constant shift of the bare mass $M_0^2$ above:
\be
\xi\,{\cal R}\,\vec \Phi^2\,\equiv\,\frac{(d-1)^2-1}{4 \,R^2}\,\vec \Phi^2\,=\,\frac{3}{4 R^2}\,\vec \Phi^2\,.
\ee
In three dimensions, the quartic coupling has mass dimension one, and therefore there are two tunable dimensionless parameters:
\be
\tilde \lambda\,=\, \lambda_0 R\,,\qquad\qquad \tilde M\,=\, M_{\rm ren}R\,.
\ee
Here $M_{\rm ren}$ denotes the renormalized or physical mass. The relation between $M_{\rm ren}$ and the bare mass $M_0$ will be made precise below. The coupling constant receives no divergent renormalization in three dimensions. We will study the time evolution of this theory as an initial value problem, by assuming that the de Sitter evolution switches on at $t=0$ and continues for all time $t>0$.  

\section{The ``E-frame" model on $S^2\times {\mathbb R}_\tau$}
Since the global de Sitter metric is conformal to  $S^2\times {\mathbb R}_\tau$ according to eq.\eqref{esumetric}, the $O(N)$ model on the fixed de Sitter background can be mapped by a conformal rescaling to a corresponding theory with time dependent couplings on $S^2\times {\mathbb R}_\tau$:
\be
\cosh \left(\tfrac{t}{R}\right)\,\equiv\, \frac{1}{\cos\tau}\,,\qquad\qquad\vec \Phi_E\,=\,\frac{\vec \Phi}{\sqrt{\cos\tau }}\,.
\ee 
We refer to this as the E-frame theory and the associated  (bare) Lagrangian density is (up to certain boundary terms),
\bea
&&{\cal L}_{{\mathbb R}_\tau\times S^2}\,=\,\frac{R^3}{2}\,\left[R^{-2}\left(\partial_\tau \vec\Phi_E\right)^2\,-\, R^{-2}\left({\bf \nabla}_{{\rm S}^2}\,\vec\Phi_E\right)^2\right.\label{ONlag}\\
&&\hspace{2in}\left.\,-\,\left(M(\tau)^2\,+\,\frac{1}{4 R^2}\right)\,\vec \Phi_E\cdot\vec\Phi_E\,-\,\frac{\lambda(\tau)}{4N}\,\left(\vec\Phi_E\cdot\vec\Phi_E\right)^2\right]\,\nonumber\\\nonumber
&& M(\tau)^2\,\equiv\,\frac{M^2_0}{\cos^2\tau}\,,\qquad\qquad\lambda(\tau)\,\equiv\,\frac{\lambda_0}{\cos \tau}\,.
\eea
When the mass and the quartic interaction are turned off, this yields a conformally coupled, free massless theory on $S^2 \times {\mathbb R}_\tau$.
The map  takes the infinite de Sitter time evolution and places it within a finite time interval in E-frame, $\tau \in \left(-\frac{\pi}{2},\,\frac{\pi}{2}\right)$.
The dS-frame and E-frame classical actions are equal provided we include the contribution from a total derivative in the E-frame action: 
\be
{\cal L}^{\rm bdry}_{\mathbb R_\tau\times S^2}\,=\,-\,\frac{R}{4}\,\frac{d}{d\tau}\left(\vec\Phi_E\cdot\vec\Phi_E\, \tan \tau\right)\,.
\ee
This term yields contributions from the initial time and the end of time at $\tau\,=\,\frac{\pi}{2}$ which are non-zero and potentially divergent  unless the E-frame fields happen to vanish sufficiently rapidly at the end of time. 

The main question that we want to address in this paper is whether the E-frame theory, with couplings diverging at finite time, has a well-defined evolution that could, in principle, be continued past the end of time at $\tau=\frac{\pi}{2}$. 

\section{The  large-$N$ limit}
The large-$N$ limit of the interacting $O(N)$ vector model on ${\mathbb R}^3$ describes a flow between the UV free fixed point and an IR  Gaussian fixed point where the operator 
\be
\sigma\,\equiv\,\frac{1}{N}\,\vec\Phi\cdot\vec\Phi\,
\ee
 acquires nontrivial scaling dimension $\Delta=2$. On ${\mathbb R}_\tau\times S^2$ or equivalently, de Sitter spacetime in global slicing, the finite spatial volume provides a spatial IR cutoff.  However, the de Sitter expansion also induces a red-shifting of the momentum modes,  which manifests itself as a time dependent coupling \eqref{ONlag} in the E-frame.  In particular, the E-frame coupling $\lambda(\tau)$ diverges at finite time, arguably forcing the theory towards  the strong coupling IR fixed point.  In order to analyze this possibility, we first implement the large-$N$ limit in the E-frame description. As usual, the  limit is implemented via a saddle-point of the path integral and turns out to be equivalent to a mean field approximation. 

An important aspect of formulating the theory in {\it global} dS$_3$  or on ${\mathbb R}_\tau\times S^2$, is that we are obliged to only focus attention on the manifestly $O(N)$-symmetric phase of the theory. Symmetry breaking is precluded in a finite spatial volume even at large $N$ (see e.g. \cite{onsquashed} for a related discussion). In order to discuss the large-$N$ limit in E-frame, we introduce the E-frame composite operator
\be
\sigma_E\,\equiv\,\frac{1}{N}\,\vec\Phi_E\cdot\vec\Phi_E\,.\label{E-framesigma}
\ee
As is standard, the bare Lagrangian for the $O(N)$ model on ${\mathbb R}_\tau\times S^2$ can be then rewritten in terms of the fundamental scalars $\Phi_E$ and the composite operator $\sigma_E$ where the relation \eqref{E-framesigma} is implemented by a Lagrange multiplier field $h$:
 \bea
 {\cal L}_{{\mathbb R}_\tau\times S^2}\,=\,\frac{R^3}{2}&&\left[\vec\Phi_E\cdot\left(-R^{-2}\partial_\tau^2\,+\,R^{-2}\nabla^2\,-\,M(\tau)^2\,-\,\frac{1}{4R^2}\,-
 \,h\right)\vec\Phi_E\right.\\\nonumber
 &&\left. -\,\frac{N\lambda(\tau)}{4}\,\sigma_E^2\,+\,Nh\sigma_E\right]\,.
 \eea
 Performing the Gaussian path integral over the elementary scalars $\vec \Phi_E$, we formally obtain (in Lorentzian signature) the large-$N$ effective action for the composite field $\sigma_E$\,,
 \bea
 &&S_{{\mathbb R}_\tau\times S^2}[\sigma_E]\,=\,i\frac{N}{2}\,{\rm Tr}\ln\left(\,R^{-2}\Box\,+\,M(\tau)^2\,+\,\frac{1}{4R^2}\,+\,h\right)\,+\\\nonumber
&&\hspace{3.0in}+\,N\int d\tau\, d^2\Omega\frac{R^3}{2}\left(h\sigma_E\,-\,\frac{\lambda(\tau)}{4}\sigma^2_E\right)\,,
 \eea
 where $\Box\equiv \partial_\tau^2\,-\,\nabla^2$.
 Extremizing with respect to the fields $h$ and $\sigma_E$, we arrive at the  large-$N$ saddle point conditions:
 \bea
 &&\langle h\rangle\,=\,\frac{\lambda(\tau)}{2}\langle\sigma_E\rangle_{\rm bare}\,\\\nonumber\\\nonumber
 && \frac{R^3}{2} \langle\sigma_E \rangle_{\rm bare}\,=\,-\frac{i}{2} \frac{1}{\rm Vol} \,{\rm Tr} \,\frac{1}{\,R^{-2}\Box\,+\,M(\tau)^2\,+\,\frac{1}{4R^2}\,+\,\langle h\rangle}\,.
 \eea
 This is the so-called gap equation for the mean field $\langle\sigma_E\rangle$, but now within a time dependent setting. In a rotationally invariant initial state, $\langle\sigma_E\rangle$ can only depend on time.
The formal saddle point conditions can be written in the form of a well-defined initial value problem. This is easy to make precise in the canonical quantization approach since the dynamics about the large-$N$ saddle point is Gaussian. In particular, we first expand the elementary fields $\vec \Phi_E$ in spherical harmonics: 
\bea
&&\Phi^{I}_E(\Omega,\, \tau)\,\equiv\,\frac{1}{\sqrt{ R}}\sum_{\ell=0}^\infty\sum_{m=-\ell}^\ell\left[a_{\ell m}^{I}\,
Y_\ell^m\left(\Omega\right)\, U_\ell(\tau)\,+\,a^{I\,\dagger}_{\ell m}\,Y_\ell^{m\,*}\left(\Omega\right)\, U_\ell^*(\tau)\right]\,,\\\nonumber\\\nonumber
&&[a_{\ell m}^{I}\,,\,a^{J\,\dagger}_{\,\ell' m'}]\,=\,\delta^{IJ}\,\delta_{\ell,\, \ell'}\,\,\delta_{m,\,m'}\,\qquad \qquad I,J\,=\,1,2 \ldots N\,.
\eea
The mode functions $U_\ell(\tau)$ satisfy the equation of motion for each harmonic labelled by the integer $\ell$,
\be
\ddot U_\ell(\tau)\,+\,\left[\left(\ell+\tfrac{1}{2}\right)^2\,
+\,R^2 M(\tau)^2\,+\,\tfrac{1}{2}\lambda(\tau) R^2\,\langle\sigma_E\rangle_{\rm bare}\right]U_\ell(\tau)\,=\,0\,,\label{eommode}
\ee
subject to a normalization condition that fixes the Wronskian according to,
\be
U_\ell(\tau)\, \dot U_\ell^*(\tau)\,-\,U_\ell^*(\tau)\, \dot U_\ell(\tau)\,=\,i\,.
\ee
It is then easy to see that each $\Phi^I_E$ and its conjugate momentum $\Pi^I\equiv R\dot\Phi^I_E$ satisfies the equal time canonical commutation relation,
\be
\left[\Phi^I_E(\tau,\,\Omega),\,\Pi^J(\tau,\,\Omega^\prime)\right]\,=\,i\,\delta^{IJ}\,\delta^{(2)}\left(\Omega\,-\,\Omega^\prime\right)\,.
\ee
Now, we can write down an explicit representation for the mean field $\langle\sigma_E\rangle_{\rm bare}$ and subsequently understand how to obtain the renormalized equations of motion. Our focus will be on an initial value problem wherein the system is  prepared in the static, equilibrium state of a free, {\em time independent} Hamiltonian for all times $\tau <0$ and the time evolution of  the coupling constants are switched on continuously at $\tau=0$.
Therefore, for all $\tau<0$, the mode functions $U_\ell(\tau)$ are simple exponentials, their normalization fixed by the condition on the Wronskian above:
\bea
&&U_\ell(\tau <0 )\,=\,\frac{e^{-iR\omega_\ell\tau}}{\sqrt{2R\omega_\ell}}\,,\qquad \omega_\ell\,=\,\sqrt{\left(\ell+\tfrac{1}{2}\right)^2 R^{-2}\,+\,M^2_{\rm ren}}\,,\\\nonumber
\\\nonumber 
&&U_\ell(0)\,=\,\frac{1}{\sqrt{2R\omega_\ell}}\,,\qquad \dot U_\ell(0)\,=\,-i\sqrt{\frac{R\omega_\ell}{2}}\,,
\eea
where $M_{\rm ren}$ is the renormalized mass at $\tau=0$, to be defined below.
The operators $\{a_{\ell m},\, a_{\ell m}^\dagger\}$ therefore refer to the oscillator states of the $\tau <0$ Hamiltonian. Taking the initial state to be an equilibrium state (vacuum or thermal) of this Hamiltonian,  the expression for the large-$N$ mean field becomes,
\bea
&&\langle\sigma_E\rangle_{\rm bare}\,=\,\frac{1}{4\pi R}\,\sum_{\ell=0}^{\ell_{\rm max}}(2\ell +1) \left|U_\ell(\tau)\right|^2\,(2 n_\ell +1)\,,\\\nonumber
&&\langle a_{\ell m}^{I\,\dagger} \,a_{\ell^\prime m^\prime}^J\rangle\,=\,n_\ell\,\delta^{IJ}\,\delta_{\ell,\ell^\prime}\,\delta_{m,m^\prime}\,,
\eea
where $\ell_{\rm max}$ is a UV cutoff on the mode number.
In an initially equilibrium thermal state the occupations numbers $n_\ell$ are given by the Bose-Einstein distribution function:
\be
n_\ell\,=\,\frac{1}{e^{\beta\omega_\ell}-1}\,.
\ee
\subsection{Renormalization}
The bare mean field contains an ultraviolet divergence as it is an expectation value of a composite operator. The nature of the divergence is clear at $\tau=0$,  when the mode functions are pure exponentials:
\be
\left.\langle\sigma_E\rangle_{\rm bare}\right|_{\tau =0}\,=\,\frac{1}{4\pi R}\sum_{\ell =0}^{\ell_{\rm max}}\frac{(2\ell +1)}{2R\omega_\ell} (2n_\ell+1)\,\sim\,\frac{\ell_{\rm max}}{4\pi R}\,.
\ee
The equation of motion \eqref{eommode}, must necessarily be finite, which means that the left hand side of the equation written in terms of bare quantities, must equal the corresponding expression written in terms of renormalised quantities:
\be
{\frac{(RM_0)^2}{\cos^2\tau}\,+\,\frac{\lambda_0}{2\cos\tau} \,\langle\sigma_E\rangle_{\rm bare}(\tau)\,=\,\frac{(RM_{\rm ren})^2}{\cos^2\tau}\,+\,\frac{\lambda_0}{2\cos\tau} \,\langle\sigma_E\rangle_{\rm ren}(\tau)\,.}
\ee 
At $\tau=0$, the (divergent)  bare mean field can be absorbed into a shift of the bare mass,
\be
M^2_{\rm ren}\,=\,M^2_0\,+\,\frac{\lambda_0 }{2}\,\left.\langle\sigma_E\rangle_{\rm bare}\right|_{\tau=0}\,,
\ee
so that the renormalized mean field is trivially vanishing at $\tau =0$:
\be
\langle\sigma_E\rangle_{\rm ren}(0)\,=\,0\,.
\ee
We should now be able to write down the renormalized quantities for all times $\tau\neq 0$ and the equations of motion in terms of these.  
The subtractions performed at $\tau=0$ essentially suffice to determine the finite, renormalized quantities for all times. However, the way this is realized is slightly subtle due to the time dependent couplings in E-frame or equivalently, the de Sitter expansion. From a  de Sitter frame perspective, the UV cutoff in momentum space must be imposed at a fixed {\em physical} momentum so that:
\be
\frac{\ell_{\rm max}}{\cosh\left(\frac{t}{R}\right)}\,=\,\ell_{\rm max}\,{\cos\tau}\,=\,\Lambda R\,,
\ee
where $\Lambda$ is a time independent UV scale which must be taken to infinity 
$(\Lambda R\to \infty)$ after necessary subtractions have been performed.
\paragraph{E-frame equations of motion:}
The time evolution of the large-$N$ theory in E-frame is thus determined by the following set of equations expressed in terms of renormalized quantities:
\bea
&&\ddot U_\ell(\tau)\,+\,\left[\left(\ell +\tfrac{1}{2}\right)^2\,+\, \frac{(M_{\rm ren} R)^2}{\cos^2\tau}\,+\,\frac{\lambda_0 R^2}{2\cos\tau}\langle\sigma_E\rangle_{\rm ren}(\tau)\right]U_\ell(\tau)\,=\,0\,,\label{esueomren}\\\nonumber\\\nonumber
&&\langle\sigma_E\rangle_{\rm ren}(\tau)\,=\,\frac{1}{4\pi R}\sum_{\ell =0}^{\Lambda R/\cos\tau}(2\ell+1) \left[\left|U_\ell(\tau)\right|^2\,{\rm coth}\left(\frac{\beta\omega_\ell}{2}\right)\,-\,\frac{1}{2R\omega_\ell}\right]\\\nonumber\\\nonumber
&& U_\ell(0)\,=\,\frac{1}{\sqrt{2R\omega_\ell}}\,,\qquad\qquad \dot U_\ell(0)\,=\,-i\sqrt{\frac{R\omega_\ell}{ 2}}\,.
\eea
This means that the number of spherical harmonic modes to be kept in the system grows with time (exponentially in de Sitter time), which is due to the red-shifting of physical momentum scales in de Sitter spacetime. In E-frame this is reflected by the unbounded growth of the time dependent mass $M^2(\tau)=M^2_{\rm ren}/\cos^2\tau$ as $\tau\to \frac{\pi}{2}$ causing all  harmonics with $\ell < M(\tau)$ to freeze out, so that $\ell_{\rm max}$ must continually be updated with time.
\paragraph{dS-frame equations of motion:} The large-$N$ saddle point conditions in dS-frame can  be directly obtained using the same steps outlined above. We can also arrive at the same, by undoing the conformal transformations that took us to the E-frame, and  introducing the de Sitter space mode functions
\be
\cos\tau\,=\,\frac{1}{\cosh\left(\frac{t}{R}\right)}\,,\qquad V_\ell(t)\,=\,\frac{U_\ell}{\sqrt{\cosh\left(\frac{t}{R}\right)}}\,, \qquad \langle\sigma_{\rm dS}\rangle\,=\,\frac{\langle\sigma_E\rangle}{\cosh \left(\frac{t}{R}\right)}\,,
\ee
obtained by solving the coupled, nonlinear system:
\bea
&&\ddot V_\ell\,+\,\tfrac{2}{R} \tanh\left(\tfrac{t}{R}\right)\,\dot V_\ell\,+\,\left(\frac{\ell(\ell+1)}{R^2\cosh^2 \left(\tfrac{t}{R}\right)}\,+\,M_{\rm ren}^2\,+\frac{3}{4 R^2}\,+\,\frac{\lambda_0}{2}\langle\sigma_{\rm dS}\rangle_{\rm ren}\right)\,V_\ell\,=\,0\,,\label{eomdsren}\nonumber\\\\\nonumber
&&\langle\sigma_{\rm dS}\rangle_{\rm ren}\,=\,\frac{1}{4\pi R}\sum_{\ell=0}^{\Lambda \cosh({t}/{R})}\,(2\ell+1)\,\left[\left|V_\ell(t)\right|^2\,\coth\left(\tfrac{\beta\omega_\ell}{2}\right)\,-\,\frac{1}{2R\omega_\ell \,\cosh \left(\frac{t}{R}\right)}\right]\,,\\\nonumber\\\nonumber\\\nonumber
&&V_\ell(0)\,=\,\frac{1}{\sqrt{2R\omega_\ell}}\,,\qquad
\dot V_\ell(0)\,=\,-i\sqrt{\frac{\omega_\ell}{2R}}\,.
\eea
\paragraph{E-frame energy:} The expectation value of the energy of the system in E-frame can also be obtained straightforwardly. Since the large-$N$ limit is equivalent to mean field theory and is Gaussian, the large-$N$ Hamiltonian (in E-frame) is
\be
H_E\,=\,\int d^2 \Omega\left[\frac{1}{2R^2}\vec\Pi^2\,+\,\tfrac{1}{2}\left(\nabla_{\rm S^2}\vec \Phi_E\right)^2\,+\, \tfrac{1}{2}\left(M^2_{\rm eff} R^2 \,+\,\tfrac{1}{4}\right) \,\vec\Phi_E\cdot\vec\Phi_E \right]
\ee 
where 
\be
\vec \Pi\,=\, R\dot{\vec \Phi}\,,\qquad \qquad M_{\rm eff}^2\,=\, \frac{M^2_{\rm ren}}{\cos^2\tau}\,+\,\frac{\lambda_0}{2 \cos\tau}\langle\sigma_E\rangle_{\rm ren}\,.
\ee
Using the expansion of the field in terms of spherical harmonics and the fact that the system is prepared in an initial equilibrium state, we find
\bea
{\cal E}&&=\,\frac{1}{N}\langle H_E\rangle\,\\\nonumber
&&=\,\sum_{\ell =0}^\infty \frac{(2\ell+1)}{2R}\left[\left| \dot U_\ell\right|^2
\,+\,\left(\left(\ell+\tfrac{1}{2}\right)^2\,+\,M^2_{\rm eff}(\tau)R^2\right)\left|U_\ell\right|^2\right] {\rm coth}\left(\tfrac{\beta\omega_\ell}{2}\right)\,.
\eea
We can use this to define the average energy per harmonic:
\be
{\cal E}\,=\,\sum_{\ell=0}^\infty (2\ell +1)\,{\cal E}_\ell\,.
\ee
An important point here is that the energy includes the zero-point fluctuations of each mode and therefore is formally divergent. Later we will be interested in the total renormalized energy.
\section{The massless theory: $M_{\rm ren}=0$}
We first discuss the theory with vanishing renormalized mass $M_{\rm ren}=0$.  In this case, we may consider two separate situations :
\begin{enumerate}
\item{$\langle\sigma_E\rangle_{\rm ren}(0)=0$: When the renormalized mass vanishes ${\rm and}$ the mean field is also vanishing in the initial state, the time evolution of the large-$N$ saddle point is trivial. It is straightforward to see that the self-consistent, renormalized gap equation is solved by 
\be
\langle\sigma_E\rangle_{\rm ren}(\tau)\,=\,0\,,\qquad U_\ell(\tau)\,=\,\frac{e^{-iR\omega_\ell \tau}}{\sqrt{2 R \omega_\ell}}\,.
\ee
Although the mean field does not evolve with time, its correlations functions do. This is one of the points we would like to address below, in order to understand whether the E-frame theory  is driven to strong coupling at late times.}
\item{$\langle\sigma_E\rangle_{\rm ren}(0)\,\neq\, 0$: This is an interesting situation wherein the initial condition (temperature) and/or the renormalization scheme is chosen such that the renormalized mean field is non-vanishing at $t=\tau=0$. In the ESU frame, its effect  on the equations of motion is to introduce a time dependent mass term $\sim 1/(\cos\tau)$ which diverges at $\tau=\frac{\pi}{2}$. Explicitly, we have (following eq.\eqref{esueomren}),
\bea
&&\frac{\lambda_0}{\cos\tau}\langle\sigma_E\rangle_{\rm ren}(\tau)\label{sigmadisplaced}\\\nonumber
&&\,=\,
\frac{\lambda_0}{\cos\tau}\left[\frac{1}{4\pi R}{\sum_{\ell =0}^{\ell_{\rm max}}}
 \left(2\ell+1\right)\left(\left|U_\ell\right|^2\,\coth\left(\tfrac{\beta\omega_\ell}{2}\right)\,-\,\frac{1}{2R\omega_\ell}\right)\,+\,\sigma_0\right]\,.
\eea
Physically, there are two ways to envisage a non-zero mean field $\langle\sigma_E\rangle$ at $\tau =0$. The first is any non-zero temperature $\beta^{-1}\neq 0$, and the second is a non-zero constant value for $\sigma_0$ due to an external source for $\sigma_E$, which is turned off at $\tau=0$ .
As we have explained earlier, $\ell_{\rm max}\simeq {\Lambda}R/\cos\tau$, and formally the cutoff $\Lambda R$ must be taken to infinity. The large-$N$ saddle point will then exhibit nontrivial evolution towards late times which we will try to understand both numerically and analytically.}
\end{enumerate}
\subsection{The massless ``critical" theory}
\label{sec:masslesscrit}
We have noted  that in the theory  with $M_{\rm ren}=0$, the large-$N$ mean field vanishes at all times if it vanishes at the initial time $\tau=0$. Despite this, fluctuations around the saddle point will exhibit nontrivial time dependence in  correlation functions of the composite field $\sigma_E$.  In order to compute the two point correlator of $\sigma_E$ about the trivial saddle $\langle\sigma_E\rangle=0$, we expand the effective action for $\sigma_E$ about this large-$N$ saddle point to quadratic order:
\bea
&& S_{\rm eff}[\sigma_E]\,=\,S_{\rm eff}[0]\,-\,i\frac{N}{16}{\rm Tr} \left[G\circ\left(\lambda\sigma_E\right)\circ G\circ\left(\lambda\sigma_E\right)\right]\,+\,N\int d\tau\,d^2\Omega \frac{R^3}{4} \lambda(\tau)\sigma_E^2\,,\nonumber\\
&&G\,=\,\frac{1}{R^{-2}\Box+\, \frac{1}{4R^2}}\,.
\eea
Since $\lambda$ depends on time, it is more convenient to define a rescaled field
\be
\tilde\sigma_E\,=\,\lambda(\tau)\sigma_E\,,\qquad\lambda(\tau)\,=\,\frac{\lambda_0}{\cos\tau}\,.
\ee
The effective action for $\tilde \sigma_E$ then has the form:
\be
S_{\rm eff}[\tilde\sigma_E]\,=\,S_{\rm eff}[0]\,-\,i\frac{N}{16}{\rm Tr} \left[G\circ\tilde\sigma_E\circ G\circ\tilde\sigma_E\right]\,+\,N\int d\tau\,d^2\Omega \frac{R^3}{4} \lambda(\tau)^{-1}\tilde\sigma_E^2\,.
\ee
As $\tau$ approaches $\frac{\pi}{2}$, the second term is vanishingly small, and the propagator for $\tilde\sigma_E$ is determined by the first term,
\be
S^{(2)}\left.\right|_{\tau \to \frac{\pi}{2}}\,=\,-i\frac{N}{16}R^6\int\, d^3x_1\int d^3 x_2
\,\tilde\sigma_E(x_1)\,G(x_1,x_2)^2\,\tilde\sigma_E(x_1)\,,\label{latetime}
\ee
where $x_{1,2}\,=\,(\tau_{1,2},\, \Omega_{1,2})$ and, crucially, $G$ is the propagator for a conformally coupled massless scalar on $S^2\times {\mathbb R}_\tau$.  This means that the late time two-point function of $\tilde\sigma_E$ can be determined from \eqref{latetime} by  conformally transforming $S^2\times {\mathbb R}_\tau$ (in Euclidean signature) to ${\mathbb R}^3$.  Performing the calculation on ${\mathbb R}^3$ by Fourier transforming to momentum space, we then obtain the result on $S^2\times {\mathbb R}_\tau$:
\be
\langle\sigma_E(\tau_1,\Omega_1)\, \sigma_E(\tau_2, \Omega_2)\rangle\left.\right|_{\tau_{1,2}\to\frac{\pi}{2}}\,\sim\,\frac{R^{-4}}{N\,(\lambda_0 R)^2} \frac{\cos(\tau_1)\,\cos(\tau_2)}{\left[\cos(\tau_1-\tau_2)\,-\,\cos(\theta_1 -\theta_2)\right]^2}\,.
\ee
 For sufficiently small proper separations, the correlator is non-vanishing as $\tau_{1,2}$ approach the end of time, and scales as expected for an operator of dimension 2.
This is the well known scaling of the two point function of $\sigma_E$ at the large-$N$ Wilson-Fisher fixed point in flat space. We  therefore conclude that the massless theory on dS$_3$ evolves to this nontrivial conformal fixed point at late times.  We note that at any other time, the theory is not conformal, since the correlator for $\tilde\sigma_E$ will involve $\lambda_0$ as an explicit scale (in addition to $R$).

\subsection{The massless theory with $\langle\sigma_E\rangle \neq 0$:}
We now consider an initial state with non-zero mean field.  As indicated above, this may arise due to a thermally excited initial state or through a classical external source for $\sigma_E$ which is turned off at $\tau =0$, and the system then allowed to evolve.

\subsubsection{Noninteracting  classical picture} 
Let us first consider the classical equations of motion, {\em without} the quantum contributions to the mean field. This means that the (quantum) interactions between the modes are set to zero. In this case the (classical) modes $U_\ell^{\rm cl}$ would obey the equations,
\be
\ddot U_\ell^{\rm cl}\,+\,\left[\left(\ell\,+\,\tfrac{1}{2}\right)^2\,+\,\frac{\lambda_0 \sigma_0 R^2}{2\cos \tau}\right]U_\ell^{\rm cl}\,=\,0\,.
\ee
This is a Schr\"odinger-like equation with a potential diverging at $\tau\,=\,\pm\frac{\pi}{2}$. In the vicinity of the singularity at $\tau=\frac{\pi}{2}$, we have two independent behaviours for any fixed $\ell$:
\bea
U_\ell^{\rm cl}\,\simeq&&A_\ell\left[\left(\tfrac{\pi}{2}-\tau\right)\,-\,\tfrac{1}{4}{\lambda_0\sigma_0 R^2}\left(\tfrac{\pi}{2}-\tau\right)^2 \,+\ldots\right] \label{masslesscl}\\\nonumber
&&\hspace{1.5in}+\, B_\ell\left[1\,-\,\tfrac{1}{2}{\lambda_0\sigma_0 R^2}\,\left(\tfrac{\pi}{2}-\tau\right)\,\ln\left(\tfrac{\pi}{2}-\tau\right)\,+\,\ldots\right]\,,
\eea
where $(A_\ell, B_\ell)$ are integration constants. Although both solutions are smooth near $\tau=\frac{\pi}{2}$, the expectation value of the energy per mode in the classical approximation (again ignoring quantum contributions to the mean field) generically diverges in this limit:
\bea
{\cal E}_{\ell}^{\rm cl}&&=\,\frac{1}{2R}\left(|\dot U_\ell^{\rm cl}|^2 \,+\, \left(\frac{\lambda_0\sigma_0 R^2}{2\cos\tau}\,+\,\left(\ell+\tfrac{1}{2}\right)^2\right)\left|U_\ell^{\rm cl}\right|^2\right) {\rm coth}\left(\tfrac{\beta\omega_\ell}{2}\right)\label{energycl}\\\nonumber\\\nonumber
&&\sim\, \frac{|B_\ell|^2}{\frac{\pi}{2}-\tau}\,.  
\eea
Although there are fine tuned initial conditions for which the classical energy per mode remains finite (those which correspond to $B_\ell =0$ at late time), generically the  divergence of the effective mass forces the E-frame energies to diverge at the classical level. We identify this as a singularity in the evolution of the classical field theory.
\subsubsection{The quantum evolution} 
Inclusion of the quantum fluctuations and self-consistently solving the large-$N$ saddle point equations reveals dramatically different 
time evolution in the full theory.
We solve the nonlinear system  \eqref{esueomren} with renormalized mean field given by eq.\eqref{sigmadisplaced}, allowing for a non-zero classical initial expectation value. Interestingly, $\sigma_0$ can also be taken negative, and so we will explore the two cases $\sigma_0 > 0$ and $\sigma_0 <0$ separately.  

\paragraph{Numerical approach:} 
Numerically, it turns out to be convenient to solve  the system \eqref{esueomren} in de Sitter time using the corresponding equations of motion \eqref{eomdsren}. The classical background $\sigma_0$ for the mean field appears in the de Sitter equations of motion via the replacement:
\be
\langle\sigma_{\rm dS}\rangle_{\rm ren}\,\to\,\langle\sigma_{\rm dS}\rangle_{\rm ren}\,+\,\frac{\sigma_0}{\cosh\left(\tfrac{t}{R}\right)}\,,\qquad\qquad
\langle\sigma_E\rangle_{\rm ren}\,=\,\frac{\langle\sigma_{\rm dS}\rangle_{\rm ren}}{\cos\tau}\,.
\ee\
In contrast to E-frame equations where $\sigma_0$ introduces a diverging effective mass, the late time contribution from $\sigma_0$ decays exponentially in de Sitter time. We have found numerical errors to be under better control in the latter setting. The de Sitter solutions can then be unambiguously transformed to E-frame.  The main drawback of the dS-frame numerical evolution is that the UV cutoff on the spherical harmonic mode number must  scale exponentially with time\footnote{In E-frame the corresponding scaling is a power law at late times, $\ell_{\rm max} \gg \left(\frac{\pi}{2}-\tau\right)^{-1}$.},
\be
\ell_{\rm max} \gg  {\rm cosh}\left(\frac{t}{R}\right)\,,
\ee
in order to avoid spurious cutoff-dependent unphysical results. In particular, this means that the number of modes we need to track increases exponentially with time and therefore limits how far we can follow the numerical evolution.   
In our numerical calculations we have used $\ell_{\max}$ between $600-800$, and carried out numerical evolution up to times  $t_{\rm max}/R \approx 4.5$ which corresponds to an ESU time $\sim \tau_{\rm max}\approx \frac{\pi}{2}-0.02$.   We also note that at each time step the value of a mode function $U_\ell$ with given $\ell$, requires knowledge of {\em all the modes} at the previous time step, due to the nonlinearity implicit in  the large-$N$ resummation.
 
\begin{figure}[h]
\centering
\includegraphics[width=3.0in]{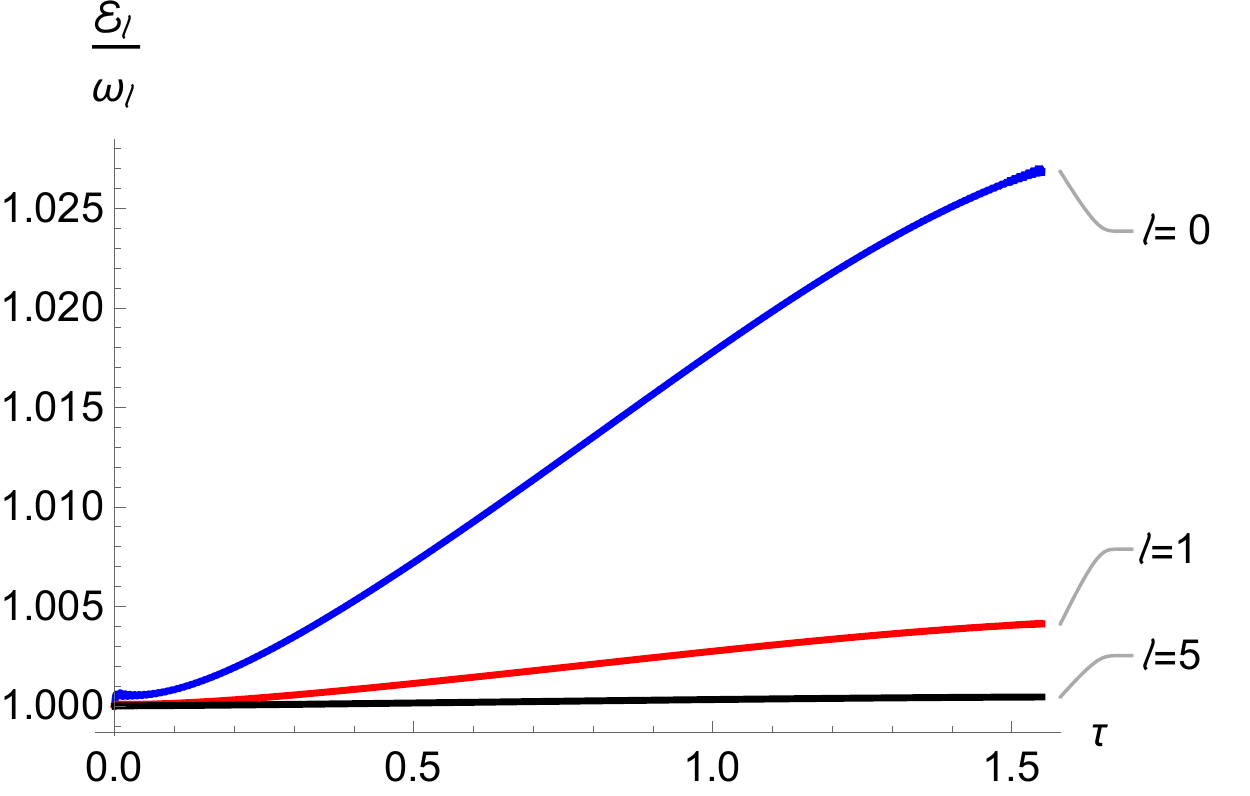}\hspace{0.05in}
\caption{\small{The energy per harmonic ${\cal E}_\ell$ divided by the initial frequency $\omega_\ell$, or the occupation number per mode as a function of $\tau$ for $\lambda_0=40\pi$ and $\lambda_0\sigma_0=0.2$.}}
\label{Tzeroenergy}
\end{figure}

\paragraph{Positive $\sigma_0$:}  The renormalized mean field has two parts, one classical and one quantum:
\be
\langle{\sigma_E}\rangle_{\rm ren}\,=\,\sigma_0\,+\,\sigma_{\rm qu}\,,\qquad
\sigma_{\rm qu}\,=\,\frac{1}{4\pi R}\sum_{\ell=0}^{\ell_{\rm max}}(2\ell +1)\left(|U_\ell|^2{\rm coth}\left(\tfrac{\beta\omega_\ell}{2}\right)\,-\,\frac{1}{2R\omega_\ell}\right)\,.
\ee
At zero temperature $\beta^{-1}=0$, and at $\tau =0$, the quantum portion $\sigma_{\rm qu}$ is vanishing. For early times and for small enough $\lambda_0$, we can ignore the effect of $\sigma_{\rm qu}$ on the evolution of the system. As the (classical) effective mass $\sim \lambda_0\sigma_0/\cos\tau$ grows with time, the magnitudes of the mode functions $|U_\ell|$ decrease, driving $\sigma_{\rm qu}$ to negative values. This has the effect of reducing  the effective mass. When backreaction from $\sigma_{\rm qu}$ becomes appreciable and the system enters the nonlinear regime, the evolution must be followed numerically. 
\begin{figure}[h]
\centering
\includegraphics[width=3.0in]{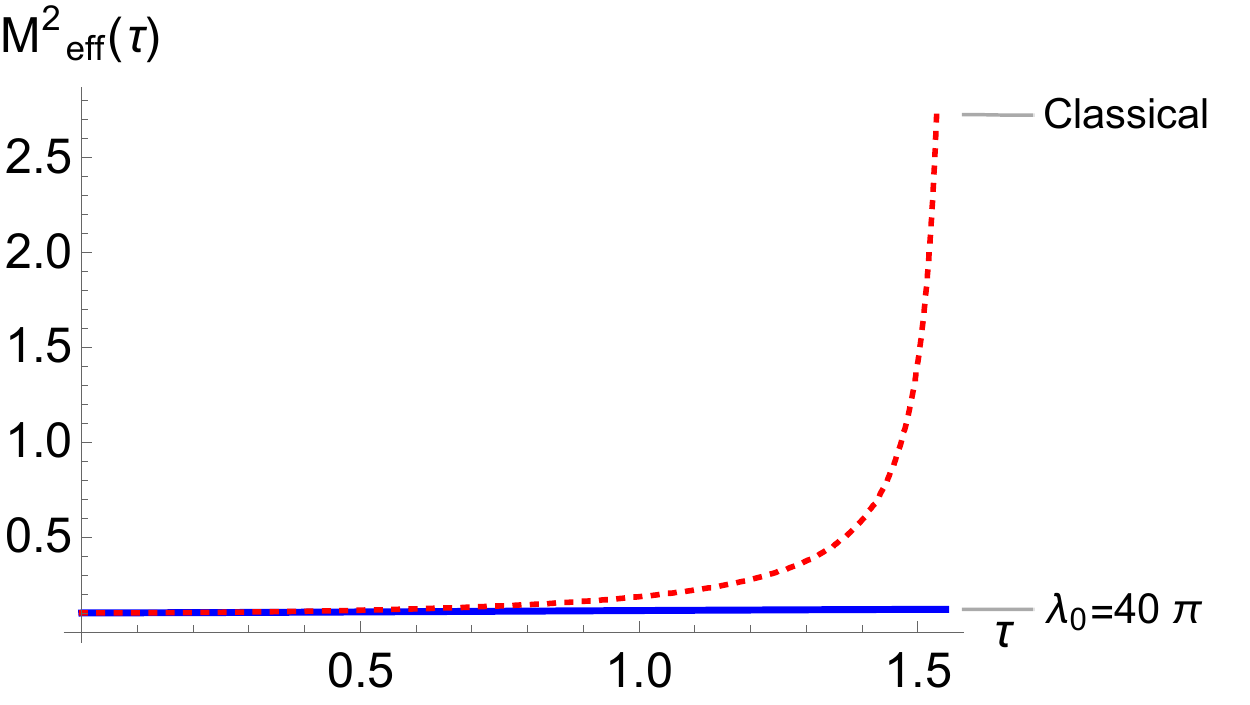}\hspace{0.05in}\includegraphics[width=2.9in]{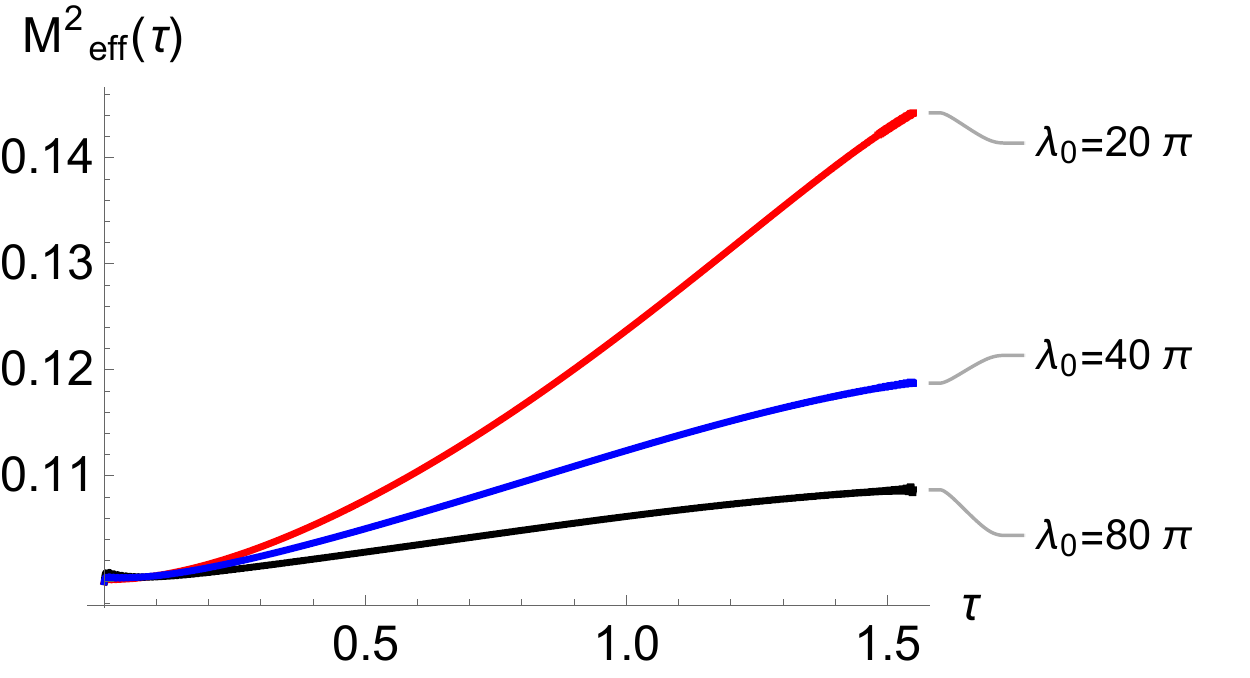}
\caption{\small{{\bf Left:} The evolution of $M^2_{\rm eff}(\tau)$ for $0 < \tau <\frac{\pi}{2}$ in the interacting theory  (solid blue) with $\lambda_0=40\pi$ and $\lambda_0\sigma_0\,=\,0.2$. For comparison we also show the classical (divergent) effective mass on the same plot. {\bf Right:} The evolution of effective mass for three different choices of $\lambda_0$ with initial condition $\lambda_0\sigma_0\,=\,0.2$.}}
\label{Tzerosigma1}
\end{figure}
Figure \ref{Tzerosigma1} shows the evolution of the effective mass:
\be
M^2_{\rm eff}(\tau)\,=\,\frac{\lambda_0}{2\cos\tau}\left(\sigma_{\rm qu}\,+\,\sigma_0\right)\,,
\ee
which remains finite as $\tau$ approaches $\frac{\pi}{2}$ even though the classical effective mass squared $=\lambda_0 \sigma_0 /(2 \cos\tau)$ diverges in this limit. This means that $\sigma_{\rm qu}$ is negative, and at late times,
\be
\left.\sigma_{\rm qu}\right|_{\tau\to \frac{\pi}{2}} \,\simeq\,- \sigma_0\,+\, \sigma_1\,\left(\tfrac{\pi}{2}-\tau\right)\,,
\ee
for some positive constant $\sigma_1$.  We can go a step further, since the numerical results show that $M^2_{\rm eff}$ flattens out and remains approximately constant for all time. This is already evident at $\lambda_0 = 80\pi$ when $M^2_{\rm eff}$ departs by less than 10\% from its initial value. 
The approximate constancy of $M^2_{\rm eff}$  at strong coupling immediately implies,
\be
\sigma_{\rm qu}\left.\right|_{\lambda_0 R\gg 1}\,\approx\,-\sigma_0\,+\,\sigma_0\,\cos\tau\,,\label{strongsigmaqu}
\ee
and the effective mass remains constant, after very early time transients. We have verified numerically that this is a very good approximation to the time dependence of $\sigma_{\rm qu}$ at strong coupling.
For general values of $\lambda_0$ the finiteness of $M^2_{\rm eff}(\tau)$ is ensured by the incorporation of the quantum fluctuations into the mean field and the value of the mass at late times is a function of the interaction strength $\lambda_0$ and the initial VEV $\sigma_0$ of the mean field,
\be
\lim_{\tau \to \frac{\pi}{2} }M^2_{\rm eff}(\tau)\,=\, M^2_{\rm final}\, \qquad\qquad M^2_{\rm final}\,=\, M^2_{\rm final}\left(\lambda_0R;\,\sigma_0 R\right)\,.
\ee
Since the effective mass squared approaches a constant at the end of time, this means that the modes $U_\ell$ behave like free field modes with fixed finite mass. Therefore, at late times,  the energy per mode also approaches a constant:
\be
{\cal E}_\ell \to \frac{1}{2R}\left[|\dot U_\ell|^2 \,+\,\left(\left(\ell+\tfrac{1}{2}\right)^2+M_{\rm final}^2 R^2\right) 
| U_\ell|^2\right]\,.
\ee
Figure \ref{Tzeroenergy} shows the evolution of the energy for a few harmonics.
Our main observation is that the energy of each mode approaches a finite constant at late time. This is important because it implies that the evolution of the modes is {\rm smooth} at ${\tau=\frac{\pi}{2}}$.
We can be more precise about the correlators of the theory at this point.  Since the masses of the modes are finite, the analysis in section \ref{sec:masslesscrit} for the two-point function of $\sigma_E$ can be repeated, with the only difference being that the full Green's function $G$ for each of the elementary scalars $\Phi^i$ is that of a massive scalar with mass $M^2_{\rm final}$  given by $G\sim \left(R^{-2}\Box\,+\,M^2_{\rm final}\right)^{-1}$, 
while the propagator for the field $\tilde \sigma_E$ at $\tau\approx\frac{\pi}{2}$ is determined by \eqref{latetime}. In the UV i.e. for proper separations smaller than $M^{-1}_{\rm final}$ this will yield the same scaling as the large-$N$ Wilson-Fisher fixed point. We can thus view the late time theory as a finite energy excited state of the Wilson-Fisher point where the fluctuations propagate with a finite effective dressed mass $M^2_{\rm final}$.

\subsubsection{Total renormalized energy}  

Although we have shown that the individual modes only acquire finite energy in the E-frame picture, we have not discussed the total energy which is formally UV divergent.  It turns out to be useful to rewrite the energy in a slightly different form using the equations of motion:
\be
{\cal E}\,=\,\frac{1}{R}\sum_{\ell=0}^{\ell_{\rm max}}(2\ell+1)\left[\frac{1}{4}\frac{d^2|U_\ell|^2}{d\tau^2}\,+\, \left(\left(\ell+\tfrac{1}{2}\right)^2\,+\,M_{\rm eff}^2(\tau)R^2\right)|U_\ell|^2\right]\coth\left(\tfrac{\beta\omega_\ell}{2}\right)\,.
\ee
We have traded the kinetic term for a total derivative using the equation of motion, and in the process, have doubled the quadratic potential contribution.
The advantage of this formula for the total energy is that the first term (after summing over $\ell$) is UV finite, and all UV divergences originate from the second term. The large $\ell$ behaviour of the summand can be deduced by assuming a WKB form for the high-$\ell$ modes:
\be
U_\ell(\tau)\,=\,\frac{e^{-i\int^\tau \varpi_\ell(\tau')}}{\sqrt{2\varpi_\ell(\tau)}}\,,\quad \qquad\varpi_\ell(\tau)\,=\,\sqrt{\left(\ell+\tfrac{1}{2}\right)^2 +M^2_{\rm eff}(\tau)R^2\,+\,\frac{3\dot\varpi^2}{4\varpi^2}-\frac{\ddot\varpi}{2\varpi}}\,.
\ee
Performing a systematic large-$\ell$ expansion, we find that
\be
\left.\frac{(2\ell+1)}{4}\frac{d^2|U_\ell|^2}{d\tau^2}\right|_{\ell \to\infty}\,\simeq\,\frac{R^2}{8\ell^2}\,\frac{d^2 M^2_{\rm eff}}{d\tau^2}\,,
\ee
which results in a UV finite sum, and
\be
\left.(2\ell+1)\left(\left(\ell+\tfrac{1}{2}\right)^2\,+\,M^2_{\rm eff}R^2\right)|U_\ell|^2\right|_{\ell \to\infty}\,\simeq\,\left(\ell +\tfrac{1}{2}\right)^2\,+\,\tfrac{1}{2}M^2_{\rm eff}R^2\,,
\ee
which leads to a UV divergent sum.  In order to define a renormalization or subtraction scheme, we first note that the bare expression for the energy depends on expectation values of composite operators:
\be
N{\cal E}\,=\,\int d^2\Omega\left[\tfrac{1}{2}\langle\dot{\vec\Phi}_E^2\rangle_{\rm bare}\,+\,\tfrac{1}{2}\left\langle\nabla_{{\rm S}^2}\vec\Phi_E\cdot\nabla_{{\rm S}^2}\vec\Phi_E\right\rangle_{\rm bare}\,+\,\tfrac{1}{2}\left(R^2M_{\rm eff}^2+\tfrac{1}{4}\right)\langle\vec\Phi_E^2\rangle_{\rm bare}\right]\,.\nonumber
\ee
The kinetic energy can be rewritten using the equations of motion, as we have done above in terms of the mode functions:
\be
\dot{{\vec\Phi}}_E^2\,=\,\frac{1}{2}\frac{d^2\, {\vec\Phi}^2_E}{d\tau^2}\,+\,\nabla_{{\rm S}^2}\vec\Phi_E\cdot\nabla_{{\rm S}^2}\vec\Phi_E\,+\,\left(R^2M_{\rm eff}^2+\tfrac{1}{4}\right)\,\vec\Phi^2_E\,.
\ee
Therefore the renormalized energy can be defined via the following renormalized expectation values:
\bea
&&\frac{1}{N}\langle\vec\Phi^2_E\rangle_{\rm ren}\,=\langle\sigma_E\rangle_{\rm ren}\,=\,\frac{1}{4\pi R}\sum_{\ell=0}^{\ell_{\rm max}}(2\ell+1)\left[|U_\ell|^2{\rm coth}\left(\tfrac{\beta\omega_\ell}{2}\right)\,-\,\frac{1}{2R\omega_\ell}\right]
\\\nonumber\\\nonumber
&&\frac{1}{N}\langle\nabla\vec\Phi_E\cdot\nabla\vec\Phi_E\rangle_{\rm ren}\,=\,\frac{1}{4\pi R}\sum_{\ell=0}^{\ell_{\rm max}}(2\ell+1)\ell(\ell+1)\left[|U_\ell|^2{\rm coth}\left(\tfrac{\beta\omega_\ell}{2}\right)\,-\,\frac{1}{2R\omega_\ell}\right]\\\nonumber\\\nonumber
&&\hspace{1.5in}-\,\frac{\ell_{\rm max}}{8\pi}\left(M^2_{\rm ren}\,-\,M^2_{\rm eff}(\tau)\right)\,.
\eea
For the finite portion of the sum, the order of the time derivatives and $\ell$-sums can be exchanged and we obtain a compact expression for the total renormalized energy:
\bea
{\cal E}_{\rm ren}\,=&&\pi\frac{d^2}{d\tau^2}\langle\sigma_E\rangle_{\rm ren}\,
-\,\frac{\ell_{\rm max}}{2}\left(M^2_{\rm ren}\,-\,M^2_{\rm eff}(\tau)\right)
\\\nonumber
&&+\,R^{-1}\sum_{\ell=0}^{\ell_{\rm max}}(2\ell+1)\left(\left(\ell+\tfrac{1}{2}\right)^2\,+\,R^2M_{\rm eff}^2(\tau)\right)\left[|U_\ell|^2\coth\left(\tfrac{\beta\omega_\ell}{2}\right)\,-\,\frac{1}{2R\omega_\ell}\right].
\eea
The cutoff $\ell_{\rm max}$ must be sufficiently large compared to 
$RM_{\rm eff}(\tau)$.  In the dS-frame, the cutoff is naturally imposed at {fixed} physical momentum $\Lambda=\ell_{\rm max}/\cosh\tau$, so that $\ell_{\rm max}$ must scale appropriately with time.  
\begin{figure}[h]
\centering
\includegraphics[width=3.0in]{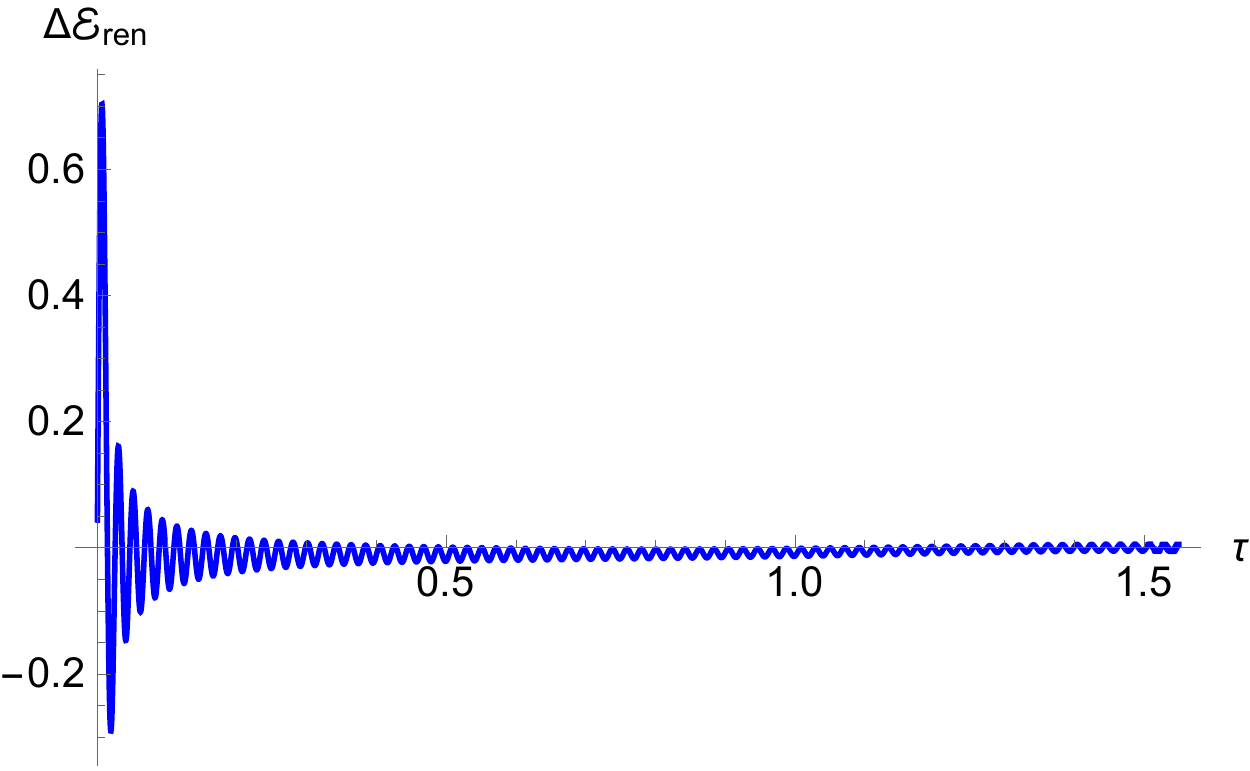}\hspace{0.05in}
\caption{\small{The change in the total renormalized energy  as a function of $\tau$ for $\lambda_0=16\pi$ and $\lambda_0\sigma_0=0.2$. The scale of the oscillations appear much larger than or comparable to the mean value.}}
\label{totenergy}
\end{figure}
The quantum evolution of the system ensures that the total renormalized energy is finite at all times since the energy deposited in each harmonic remains finite. This is shown explicitly in figure \ref{totenergy} for the theory with $\lambda_0=16\pi$ and 
$\lambda_0\sigma_0\,=\,0.2$. The numerical results show significant oscillations on short time scales whose origin is in the discretization errors inherent in the time evolution.

\subsubsection{Negative $\sigma_0$} 
It is interesting to analyze a situation where $\sigma_0$ is {\em negative}.  The effective mass-squared is then initially negative. In this case we characterize the initial value problem as follows. We define $\sigma_0(\tau)$ as:
\be
\sigma_0(\tau)\,=\, |\sigma_0|\,\Theta(-\tau)\,-\,|\sigma_0|\,\Theta(\tau)\,.
\ee
This choice is necessary in order to have a well defined initial value problem, so that the system is prepared in an equilibrium, stable state for $\tau<0$. In particular, the initial state corresponds to a free field theory with  a positive effective mass-squared for all fluctuations:
\be
M_{\rm eff}^2(\tau<0)\,=\,\frac{1}{2}\lambda_0|\sigma_0|\,.
\ee
At $\tau=0$, the time dependence in the couplings is switched on, along with a simultaneous sign flip in $\sigma_0$:
\be
M^2_{\rm eff}(\tau>0)\,=\,\frac{\lambda_0}{2\cos\tau}\left(-|\sigma_0|\,+\,\sigma_{\rm qu}\right)\,.
\ee
Without any interactions, or incorporation of quantum fluctuations, a large enough negative (growing) mass-squared would lead to growth of the modes at early times. 
A point to note is that even in the non-interacting theory (i.e. $\sigma_{\rm qu}=0$), the modes do not grow without bound. This is due to the $1/\cos\tau$ time dependence of the negative mass-squared. As in the positive $\sigma_0$ situation, it leads to regular behaviour \eqref{masslesscl} for classical solutions at $\tau=\frac{\pi}{2}$. However, if the modes followed classical behaviour at all times, the energy per mode would diverge (to negative infinity) as eq.\eqref{energycl}.  

In the interacting theory, the initial growth of quantum fluctuations backreacts on the system through the mean field $\sigma_{\rm qu}$.  It is {\em a priori} not obvious that this growth in quantum fluctuations will be sufficient to cancel the $(1/\cos\tau)$ dependence of the classical, negative mass-squared.
When the initial state is prepared at zero temperature, $\beta^{-1}=0$, the renormalized mean field $\sigma_{\rm qu}$ is vanishing.  The equation of motion for any given mode, 
\be
\ddot U_\ell\,+\,\left[\left(\ell+\tfrac{1}{2}\right)^2\,+\,M^2_{\rm eff}(\tau) R^2\right]U_\ell\,=\,0\,,
\ee
suggests that when $M_{\rm eff}^2$ is sufficiently negative for a fixed (low) $\ell$,  at early times the mode will grow. This in turn will cause the quantum contribution $\sigma_{\rm qu}$ to the mean field to grow to positive  values.  Whether the growth is sufficient to offset the time dependence of the (negative) classical piece in $M^2_{\rm eff}$, can be understood numerically.
 
\begin{figure}[h]
\centering
\includegraphics[width=2.8in]{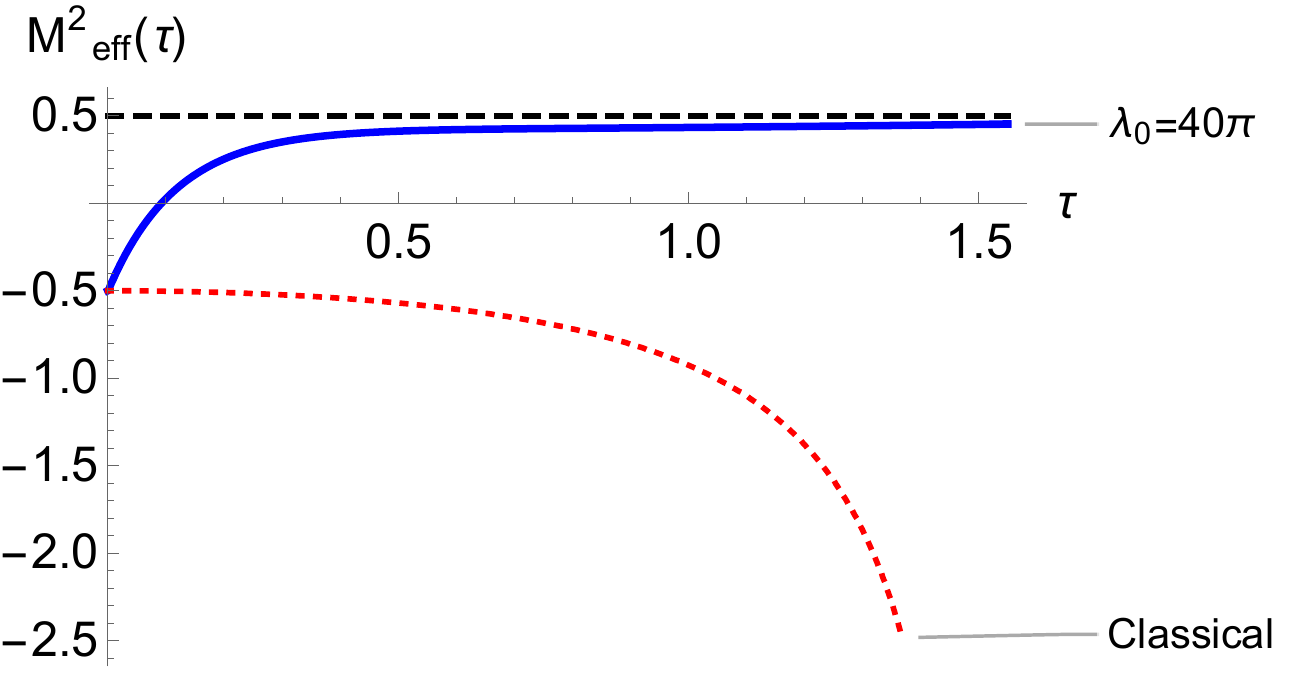}\hspace{0.2in}\includegraphics[width=2.5in]{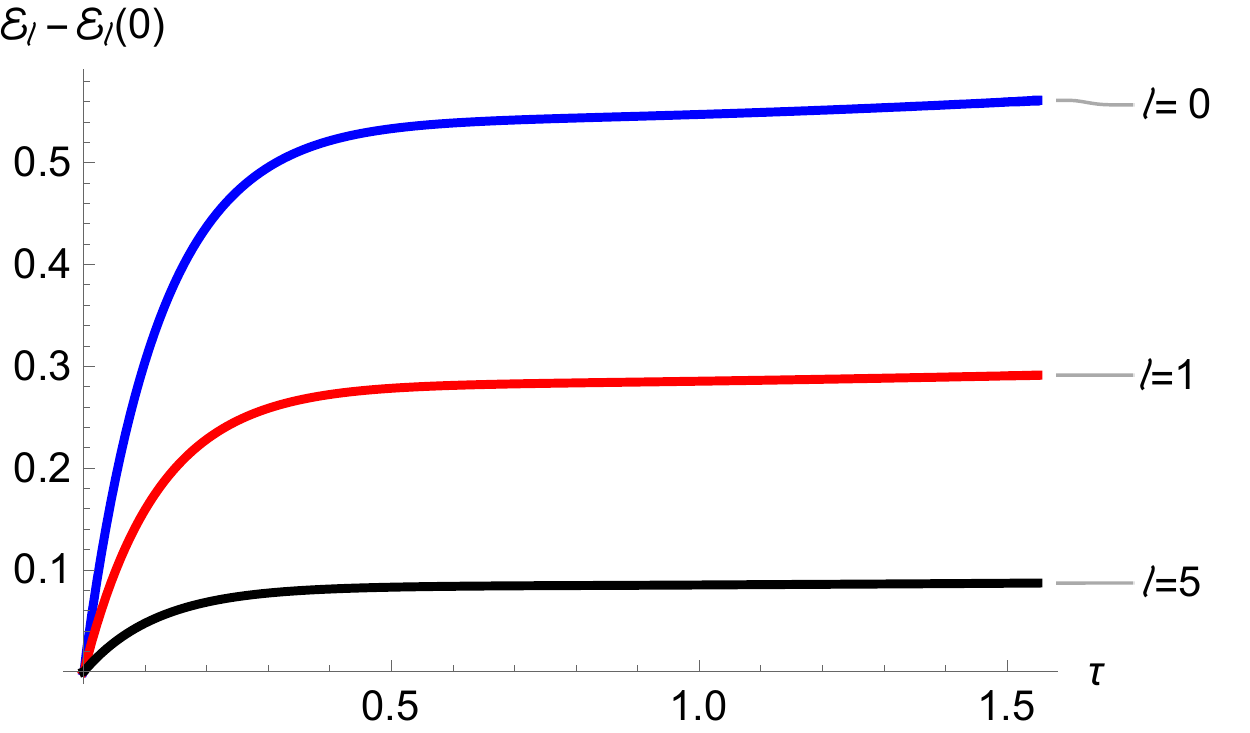}
\caption{\small{{\bf Left:} The effective mass squared starts off at a negative value ($=-0.5$) at $\tau=0$, but rapidly saturates to a positive value $\approx 0.5$ at large values of the coupling $\lambda_0=40\pi$. {\bf Right:} For the same initial conditions are shown the evolution of energy per mode for three different $\ell$ harmonics.}}
\label{negativesigma}
\end{figure}
Figure \ref{negativesigma} shows the numerical results for $M^2_{\rm eff}(\tau)$ with $\lambda_0 = 40 \pi$ and $\lambda_0|\sigma_0|\,=\,1$, for which the $\ell=0$ harmonic starts off unstable at $\tau=0$. Due to the large value of the coupling, we see rapid growth of $M^2_{\rm eff}$ to positive values saturating at $M^2_{\rm eff}\approx0.5$:
\be
\sigma_{\rm qu}\left.\right|_{\tau\to\frac{\pi}{2}}\,\approx\,|\sigma_0|(1\,+\,\rm \left(\tfrac{\pi}{2}-\tau\right))\,.
\ee
In all cases we see a finite amount of energy deposited in each mode, consistent with a smooth evolution of the system towards the end of time at  $\tau=\frac{\pi}{2}$.

A noteworthy point here is that although the effective mass starts off negative, there is no symmetry breaking and consequently there are no Goldstone bosons as the system is in a finite volume.
\subsection{Discussion and comparison with classical theory}
\begin{figure}[h]
\centering
\includegraphics[width=2.5in]{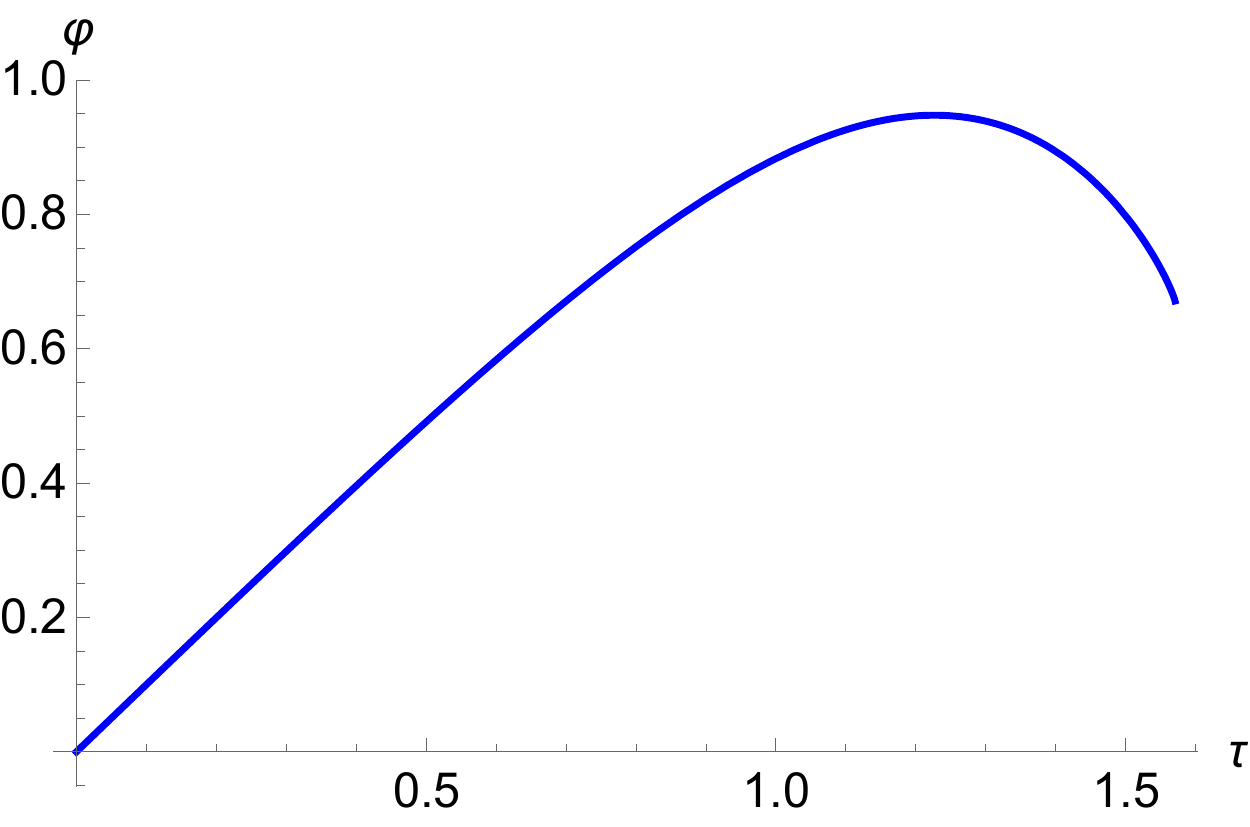}\hspace{0.5in}\includegraphics[width=2.5in]{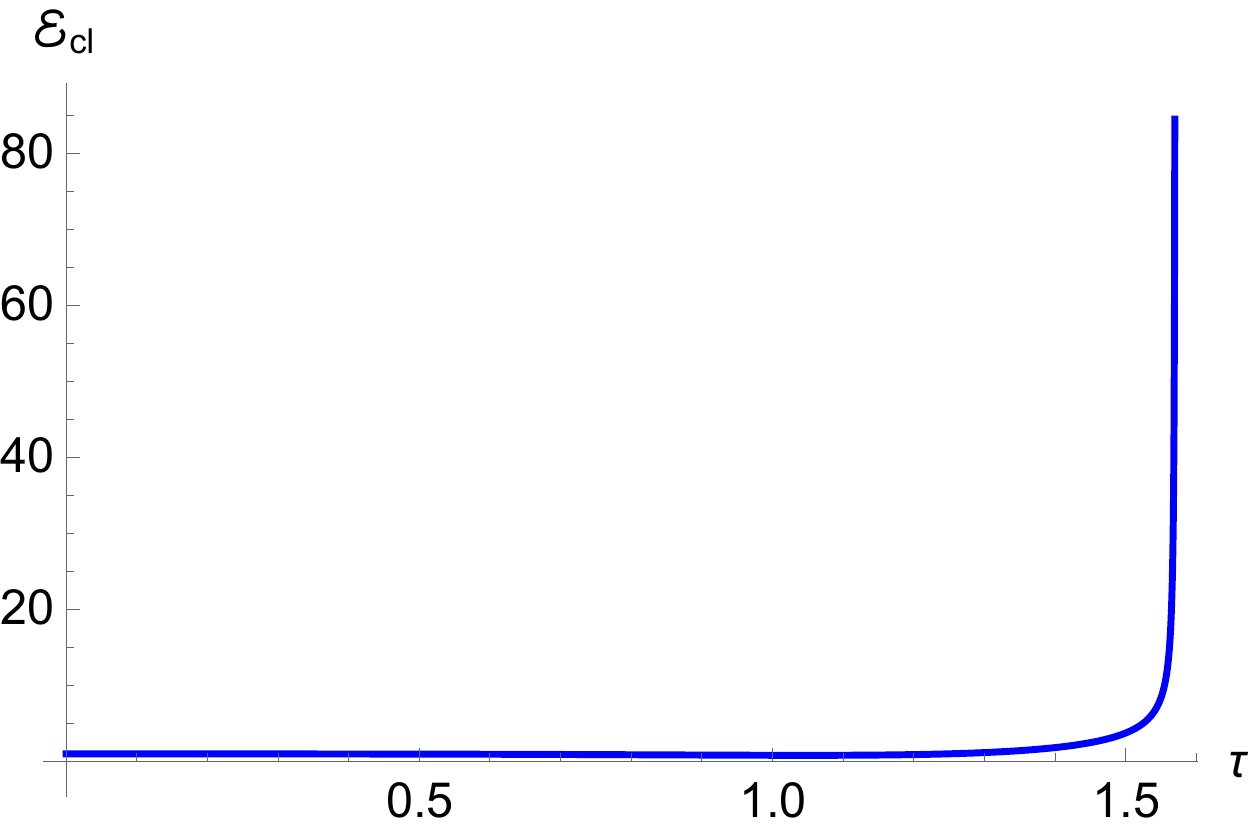}
\caption{\small{The classical field evolves smoothly towards $\tau=\frac{\pi}{2}$ whilst the energy diverges at this time. The trajectory shown is for $\sigma_0\,=\,0.1$, $\lambda_0=1$, $\varphi(0)=0$ and $\varphi'(0)=1$.}}
\label{classical}
\end{figure}
The results of the quantum large-$N$ evolution of the system are dramatically different from pure classical field theory even after interactions are switched on in the classical system. 
To make the comparison explicit, let us consider the classical interacting theory and in particular, focus on the classical dynamics of the zero mode $\varphi(\tau)$ of one of the $N$-tuplet of scalars in E-frame (setting all other fields to zero consistently, and performing a rescaling $\varphi\to\sqrt N\varphi$):
\be
{\cal L}_{\rm cl}\,=\,\frac{1}{2R^2}\left(\dot \varphi^2\,-\,\frac{1}{4}\varphi^2\,-\,\frac{\lambda_0\sigma_0 R^2}{2\cos\tau}\varphi^2\,-\,\frac{\lambda_0 R^2}{4\cos\tau}\,\varphi^4\right)\,.
\ee
The equation of motion for the driven system,
\be
\ddot\varphi\,+\,\left(\frac{1}{4}\,+\,\frac{\lambda_0 R^2}{2\cos\tau }\left(\sigma_0+ \varphi^2\right)\right)\varphi\,=\,0\,,
\ee
describes the motion of a particle in a time dependent quartic potential. The evolution of $\varphi(\tau)$ is smooth, approaching a constant at $\tau=\frac{\pi}{2}$. The energy, on the other hand, diverges  (figure \ref{classical}) as $\left(\frac{\pi}{2}-\tau\right)^{-1}$. When $\sigma_0$ is negative, we have an essentially identical situation. In this case the energy diverges generically to either positive or negative infinity, and for a sufficiently fine-tuned initial condition it can  approach a finite value at $\tau=\frac{\pi}{2}$. 

In the quantum (large-$N$) theory, $\varphi^2$ is replaced by the quantum mean field $\sigma_{\rm qu}$ which receives contributions from an infinite number of degrees of freedom. The coupled mode equations then show that quantum fluctuations {\em decrease} from their initial value at $\tau=0$, and the decrease is sufficiently rapid (see e.g. eq.\eqref{strongsigmaqu}) to offset the putative divergence in the classical couplings, resulting in a finite effective mass-squared $M_{\rm eff}^2(\tau)$.  This ensures that the energy per mode remains finite  at the end of the time evolution. For negative $\sigma_0$, the quantum contributions {\em increase} rapidly with time and saturate to the positive value given by $|\sigma_0|$, yielding a finite positive $M^2_{\rm eff}(\tau)$ at late time. In particular, this endpoint is a self-consistent solution to the field equations.

\section{The massive case: $M_{\rm ren}\neq 0$}
The evolution of the system with a non-zero bare  mass can be expected to be very different to the critical case. This is because the effective mass squared has a classical piece that diverges quadratically as $\tau$ approaches $\tfrac{\pi}{2}$. The classical wave equation with $\langle\sigma_E\rangle_{\rm ren}$ set to zero (or noninteracting situation) is,
\be
\ddot U_\ell^{\rm cl}(\tau)\,+\,\left[\left(\ell+\tfrac{1}{2}\right)^2\,+\,\frac{M^2_0 R^2}{\cos^2\tau}\right]\,U_\ell^{\rm cl}(\tau)\,=\,0\,.
\ee
The free mode equation can be solved analytically in terms of Legendre polynomials.
The divergence of the Schr\"odinger potential forces solutions $U_\ell^{\rm cl}(\tau)$ for any finite $\ell$ to display non-analytic behaviour at late times,
\be
U_\ell^{\rm cl}(\tau)\,\to\,\sqrt{\tfrac{\pi}{2}-\tau}\left(A_\ell\,\left(\tfrac{\pi}{2}-\tau\right)^{i \delta}\,+\,B_\ell \,\left(\tfrac{\pi}{2}-\tau\right)^{-i\delta}\right)\,,\qquad\delta\,=\,\sqrt{M^2_{0}R^2-\tfrac{1}{4}}\,.
\ee
Therefore when $M_0^2R^2 > \tfrac{1}{4}$, all modes vanish at the end of time
whilst the energy in the E-frame (for each harmonic) diverges as
\be
{\cal E}^{\rm cl}_\ell\,\sim\,\frac{1}{\left(\tfrac{\pi}{2}-\tau\right)}\,.
\ee
For small masses $M^2_0R^2 <\tfrac{1}{4}$, the modes continue to vanish and the energy diverges, with a different exponent.
The immediate question then is whether the quantum large-$N$ dynamics can cure this singularity. The effective mass-squared in E-frame,
\be
M_{\rm eff}^2\,=\, \frac{M^2_{\rm ren}}{\cos^2\tau}\,+\,\frac{\lambda_0}{2\cos\tau}\langle\sigma_E\rangle_{\rm ren}\,,\label{fullmeff}
\ee
can reach a finite value only if $\langle\sigma_E\rangle_{\rm ren} \to - 2M^2_{\rm ren}/(\lambda_0\cos\tau)$ as $\tau\to\frac{\pi}{2}$.
This means that $\langle\sigma_E\rangle_{\rm ren}$ must not only be negative, but must diverge.  Before turning to the numerics, let us explain how such a cancellation could potentially occur.

At any  $\tau$ approaching $\frac{\pi}{2}$, working at zero temperature, we may separate out the contributions to the mean field, into a `UV' and an `IR' piece:
\bea
&&\langle\sigma_E\rangle_{\rm ren}\,=\, \sigma_{\rm UV}\,+\,\sigma_{\rm IR}\,,\qquad
\sigma_{\rm IR}\,=\,\frac{1}{4\pi R}\sum_{\ell=0}^{[M_{\rm ren}R/\cos\tau]}(2\ell+1)\left[|U_\ell|^2\,-\,\frac{1}{2R\omega_\ell}\right]\,\\\nonumber\\\nonumber
&&
\sigma_{\rm UV}\,=\,\frac{1}{4\pi R}\sum_{[RM_{\rm ren}/\cos\tau]}^{\infty}(2\ell+1)\left[|U_\ell|^2\,-\,\frac{1}{2R\omega_\ell}\right]\,.
\eea
We focus attention on the weakly coupled theory $(\lambda_0 R\ll 1 )$, so we can
work with the free (classical) modes.  We also assume $M_{\rm ren}R > \tfrac{1}{2}$.
This means that as $\tau \to \frac{\pi}{2}$, the IR modes vanish as $|U_\ell| \sim \sqrt{\frac{\pi}{2}-\tau}$, while most UV modes (large enough $\ell$) that are oscillatory will continue to have $|U_\ell| \approx \frac{1}{2\ell}$. 
More carefully, using a WKB approximation for the large $\ell$ modes we arrive at the estimates,
\be
\sigma_{\rm IR}\,\approx\, \,-\,\frac{M_{\rm ren}}{4\pi\cos\tau}\,,\qquad
\sigma_{\rm UV}\,\approx\, \frac{M_{\rm ren}}{8\pi \cos\tau}\,, \qquad
\langle\sigma_E\rangle_{\rm ren}\,\approx\, -\frac{M_{\rm ren}}{8\pi \cos\tau}\,,
\ee
which are valid as we approach the end of time $\tau\to\frac{\pi}{2}$.  The main point to take away here is that (for positive $M^2_{\rm ren}$) the quantum mean field goes negative and grows at the rate required to negate the effect of the classically divergent $M_{\rm eff}^2$ in eq.\eqref{fullmeff}. However, it is also clear that a complete cancellation will not occur when $\lambda_0$ is small. The remaining question is whether such cancellation can occur as the coupling is increased. We analyze this possibility below numerically.

\subsection{Large-$N$ dynamics with $M_{\rm ren} \neq 0$}
We first recall that the quantum mean field in de Sitter and E-frames are related as,
\be
\langle\sigma_E\rangle\,=\,\frac{\langle\sigma_{\rm dS}\rangle}{\cos\tau}\,,
\ee 
and the effective mass \eqref{fullmeff} is,
\be
M^2_{\rm eff}\,=\,\frac{1}{\cos^2\tau}\left(M_{\rm ren}^2\,+\,\frac{\lambda_0}{2}\langle\sigma_{\rm dS}\rangle_{\rm ren}\right)\,.
\ee
If the de Sitter mean field approaches a (negative) constant  at late times, we expect on general grounds that as $\tau\to\frac{\pi}{2}$, assuming a regular power series expansion,
\be
\langle\sigma_{\rm dS}\rangle_{\rm ren} \,=\, \sum_{n=0}^\infty s_n\left(\tfrac{\pi}{2}-\tau\right)^n\,,\qquad \cos\tau\,=\,\frac{1}{\cosh\left(\frac{t}{R}\right)}\,.
\ee
For smooth evolution  in E-frame at late times we  require $\lambda_0 s_0=-2M_{\rm ren}^2$ {\em and} $s_1=0$. We have seen already that the theory with  small $\lambda_0$ does not exhibit a complete cancellation of the leading late time divergence in $M^2_{\rm eff}$. In order to explore this numerically and for arbitrary couplings, it is best to work with dS-frame equations of motion \eqref{eomdsren} and examine the effective mass-squared in dS$_3$:
\be
M^2_{\rm dS}\,=\,M^2_{\rm ren}\,+\,\frac{\lambda_0}{2}\langle\sigma_{\rm dS}\rangle_{\rm ren}\,.\label{dsmass}
\ee
\subsubsection{Positive $M^2_{\rm ren}$}
\begin{figure}[h]
\centering
\includegraphics[width=2.9in]{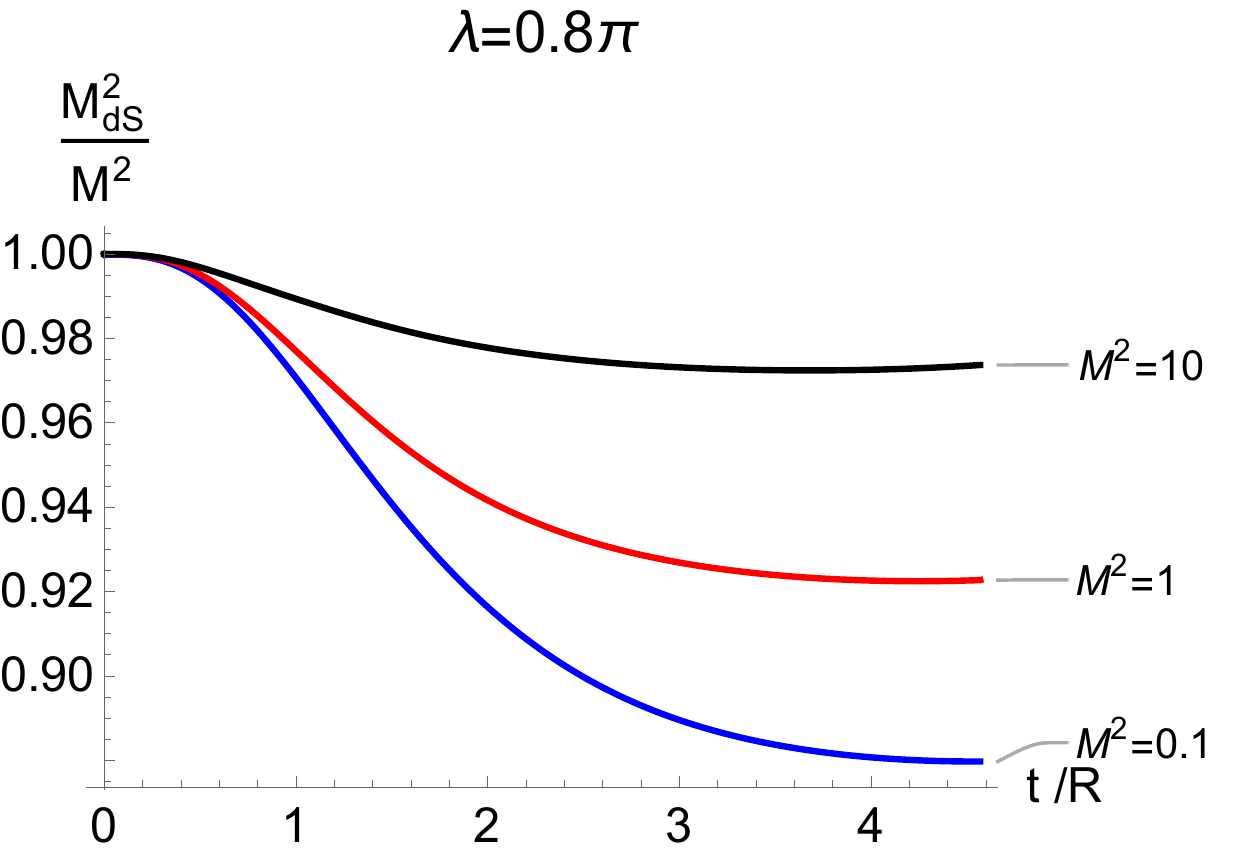}
\hspace{0.05in}\includegraphics[width=2.9in]{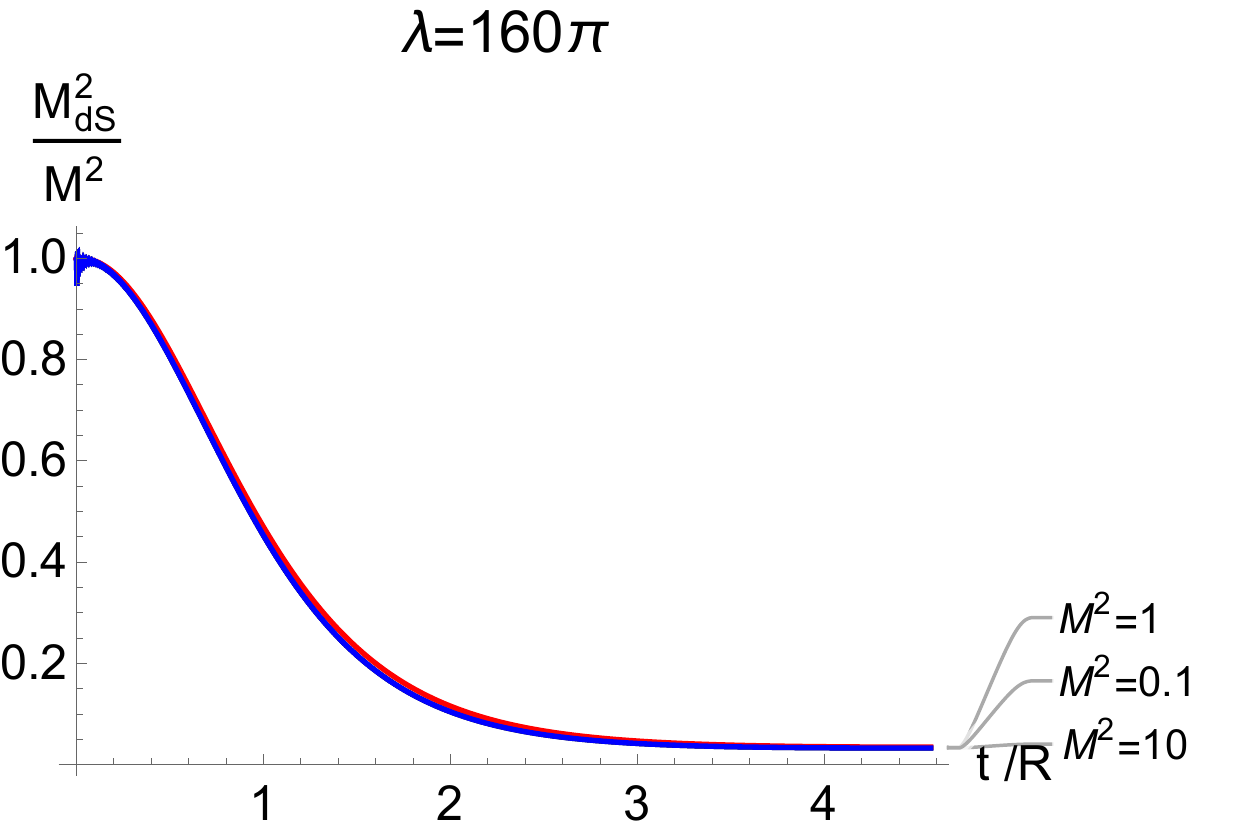}
\caption{\small{The effective mass in dS-frame as a function of global de Sitter time for three different values of the renormalized mass parameter $M^2_{\rm ren}=0.1, 1, 10$, and for weak (left) and strong (right) coupling. At strong coupling, the 3 distinct curves for $M_{\rm dS}^2/M^2_{\rm ren}$ lie on on top of each other.}
}
\label{msqds}
\end{figure}

We now present the numerical solutions to the massive saddle point equations in dS-frame.  For simplicity we set the initial temperature $\beta^{-1}=0$. We keep a total of $\ell_{\rm max}=600$ modes, and evolve from $t=0$ to $t/R \approx 4.5$.  Numerically, the upper time limit is approximately when the effects of the finite cutoff start kicking in and affecting the results.
We find that the effective mass \eqref{dsmass} in dS-frame attains a constant asymptotic  value $M_{\infty}$ such that 
\be
\lim_{t\to\infty} M_{\rm dS}(t)\,=\,M_{\infty}\,,\qquad M_{\infty} < M_{\rm ren}\,.
\ee
We also observe (figure \ref{msqds}) that as the coupling $\lambda_0 R$ is increased, the asymptotic value $M^2_\infty$ decreases monotonically. Remarkably, in the large-$\lambda_0$ regime, the time evolution displays a universality, independent of $M^2_{\rm ren}$.

We can draw certain useful inferences from this behaviour at strong coupling. In particular, the late time asymptotics in dS-frame is consistent with that of a {\em free field} theory with fixed mass $M_\infty^2$.  In the theory with large $\lambda_0 R$, we  find $M_\infty^2 R^2 \ll 1$.

We know from standard free field results in de Sitter spacetime \cite{Birrell:1982ix, dowker} that in the limit of small masses the quantum mean field  in the Bunch-Davies vacuum state depends quadratically on the mass:
\be
\langle\sigma_{\rm dS}\rangle\left.\right|_{\rm Bunch-Davies}\,\simeq\,-\tfrac{1}{8}M^2_{\rm ren}R\,+\,{\cal O}(M^4_{\rm ren}R^3)\,,\qquad M_{\rm ren}R \ll 1\,,
\ee
In the large mass limit the Bunch-Davies  mean field yields linear dependence on the mass: $\langle\sigma_{\rm dS}\rangle \simeq -M_{\rm ren}/4\pi$, independent of $R$. 

\begin{figure}[h]
\centering
\includegraphics[width=2.9in]{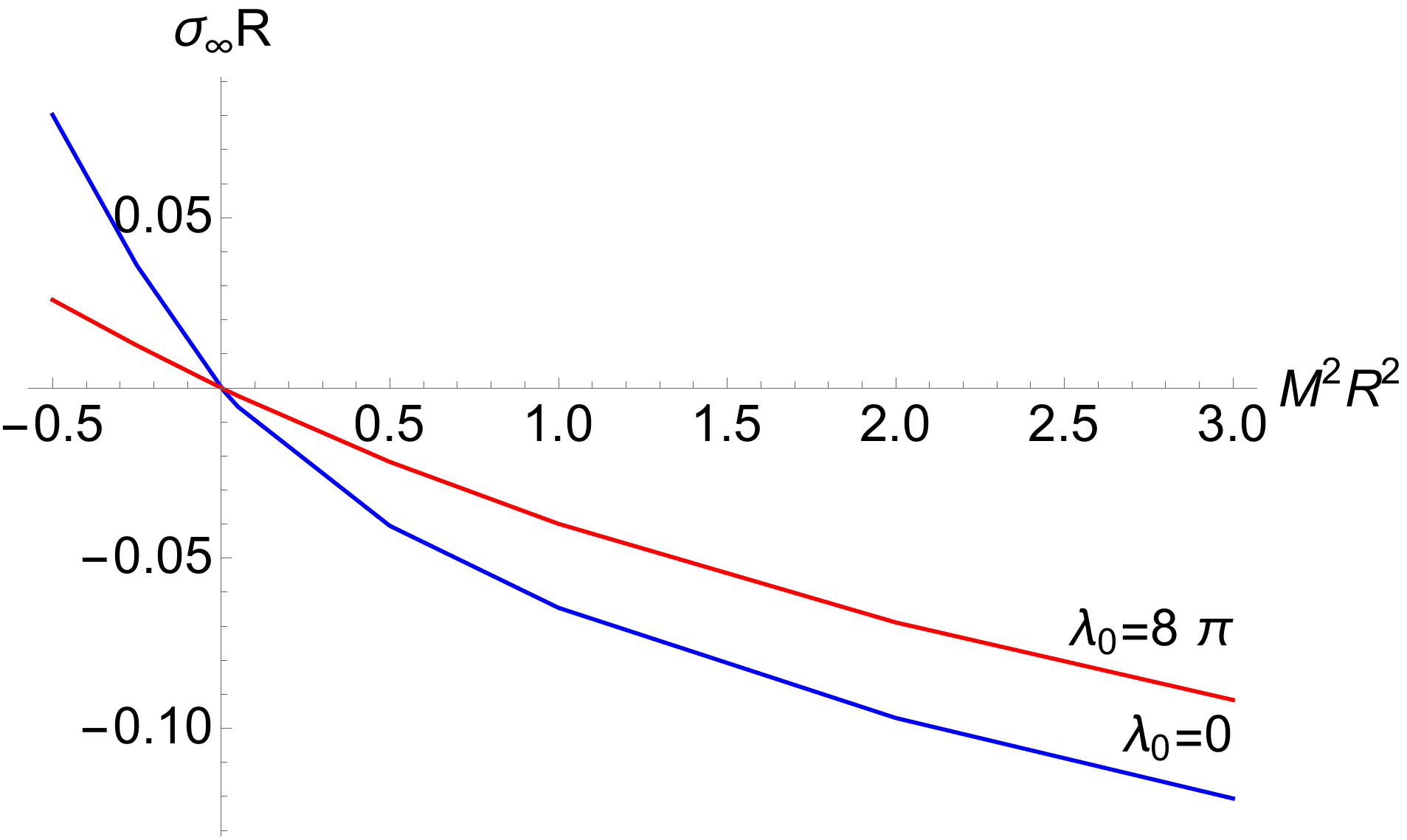}
\hspace{0.1in}\includegraphics[width=2.9in]{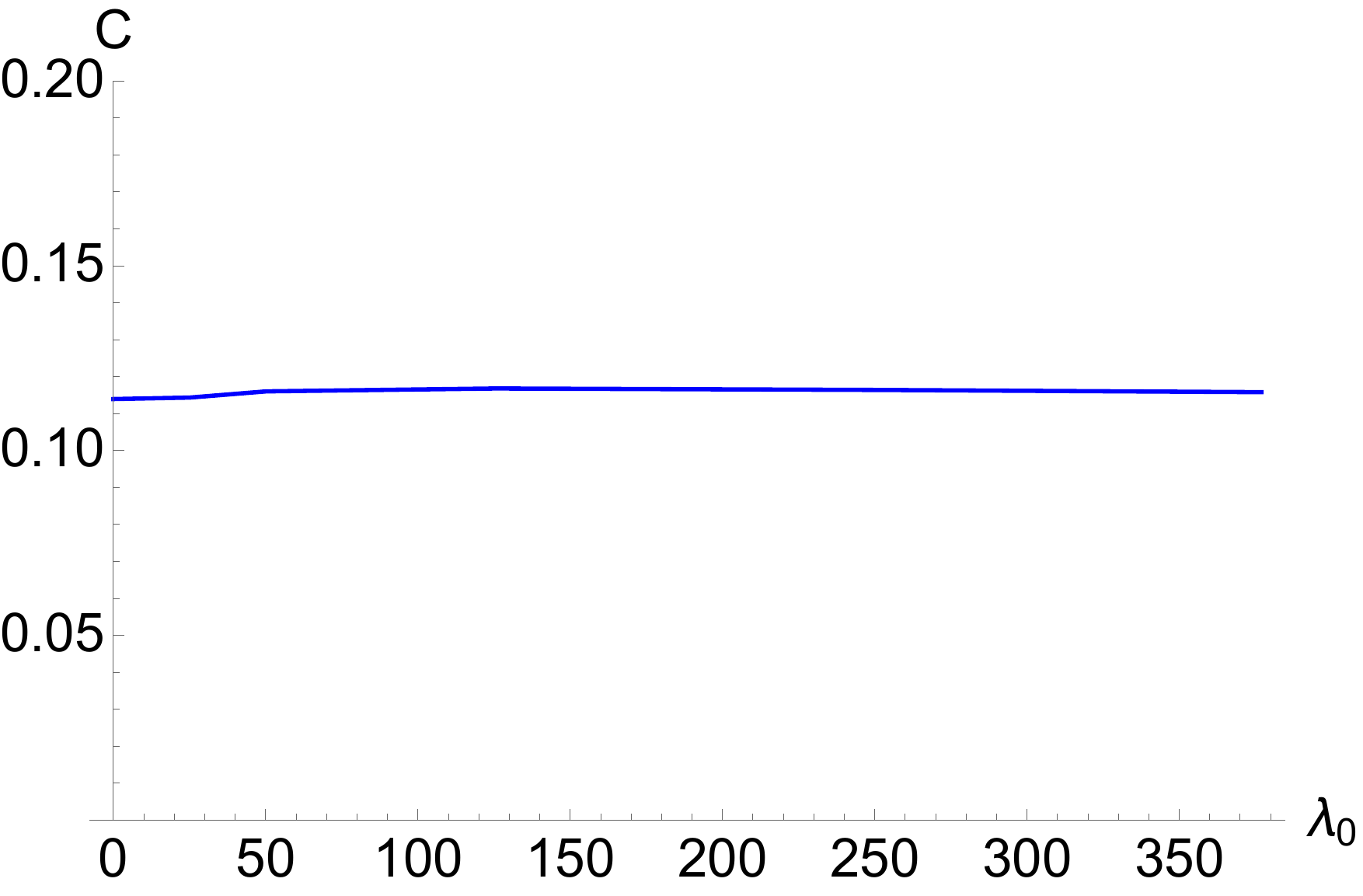}
\caption{\small{{\bf Left}: $\sigma_\infty$ as a function of $M^2
_{\rm ren}$ for both the free and interacting theories. In both cases we identify linear behaviour at small $M_{\rm ren}$. {\bf Right:} ${\cal C}(\lambda_0)$ as a function of $\lambda_0$ is almost a constant and close to the Bunch-Davies value $1/8 =0.125$.  The plot obtained was for $M_{\rm ren}^2R^2=0.1$.}}
\label{sigmavsmsq}
\end{figure}

In our case, although the interacting theory relaxes to an effective, free massive 
theory on dS$_3$ we do not know {\em a priori}, whether the state at late times approaches the Bunch-Davies vacuum. Extracting the asymptotic value of $\langle\sigma_{\rm dS}\rangle_{\rm ren} $ for different values of $M^2_{\rm ren}$ 
in the free theory with the equilibrium initial conditions, we find (figure \ref{sigmavsmsq}) that $\langle\sigma_{\rm dS}\rangle$ is negative and proportional to $M_{\rm ren}^2$ for small mass.

From the observations made above we conclude that at strong coupling, when the late time asymptotics reduces to that of a free theory with a small effective mass, the asymptotic value of the dS-frame mean field takes the form,
\be
\lim_{t\to\infty}\langle\sigma_{\rm dS}\rangle_{\rm ren}\,=\,\sigma_\infty\,=\,-{\cal C}\, M_{ \infty}^2 R\,+\,{\cal O}(M^4_{ \infty})\,\qquad {\lambda_0 R\gg 1}\,.
\label{sigmainf}
\ee
for ${\cal C}$ a positive constant, potentially depending on $\lambda_0$.
Then, applying the definition of the dS-frame effective mass \eqref{dsmass} at late times, we can write,
\be
M_\infty^2\,=\,\frac{M_{\rm ren}^2}{1+\frac{1}{2}\lambda_0 R\,{\cal C}} \,\simeq\, \frac{2 M_{\rm ren}^2}{\lambda_0 R\,{\cal C}}\,,\label{minf}
\ee
where the second step assumes that ${\cal C}$ is either a monotonic function of $\lambda_0$ or a constant at large $\lambda_0$.
Extracting ${\cal C}$ for different values of the interaction strength (figure \ref{sigmavsmsq}) for small values of $M^2_{\rm ren}$,  we find that ${\cal C} \approx 0.12$, independent of $\lambda_0$. Furthermore, using eqs. \eqref{sigmainf} and \eqref{minf}, we find that 
\be
\sigma_\infty\,\simeq\,- \frac{2M^2_{\rm ren}}{\lambda_0}\,,
\ee
which is simply the condition for vanishing effective mass in de Sitter spacetime.
\subsubsection{Negative $M^2_{\rm ren}$}
The theory with negative mass-squared is ill-defined  when the interaction strength $\lambda_0$ is vanishing and the mass squared is sufficiently negative i.e. when
\be
R^2M_{\rm ren}^2 < -\frac{3}{4} 
\ee
in three dimensional de Sitter spacetime.  In this regime, free field modes in dS$_3$ {\em grow} exponentially with time. 
When interactions $\lambda_0>0$ are switched on, the classical potential has a minimum away from the origin. Recalling that finite volume precludes the breaking of the global $O(N)$-symmetry, we consider the time evolution of the system prepared in the $O(N)$-symmetric equilibrium initial state, where the interactions and time evolution are switched on at $t=0$:
\bea
&&{\text {dS-frame:}} \quad M^2(t)\,\,=\,\,|M_{\rm ren}|^2\,\Theta(-t)\,\,+\,\,M_{\rm ren}^2\,\,\Theta(t)\,,\\\nonumber\\\nonumber
&& {\text {E-frame:}} \quad M^2(\tau)\,\,=\,\,|M_{\rm ren}|^2\,\Theta(-\tau)\,\,+\,\,\frac{M_{\rm ren}^2}{\cos^2\tau}\,\Theta(\tau)\,.
\eea
This is an example of a quench where the mass-squared is rapidly driven to negative values at $t=0$, and the system finds itself in an unstable configuration at the top of the ``mexican hat" potential. For small $\lambda_0$ at least, we expect that the instability leads to an exponential growth in the modes $V_\ell$, which eventually feed into the growth of the mean field $\langle\sigma_{\rm dS}\rangle_{\rm ren}$ and thus backreacts on the large-$N$ dynamics. 
\begin{figure}[h]
\centering
\includegraphics[width=3.5in]{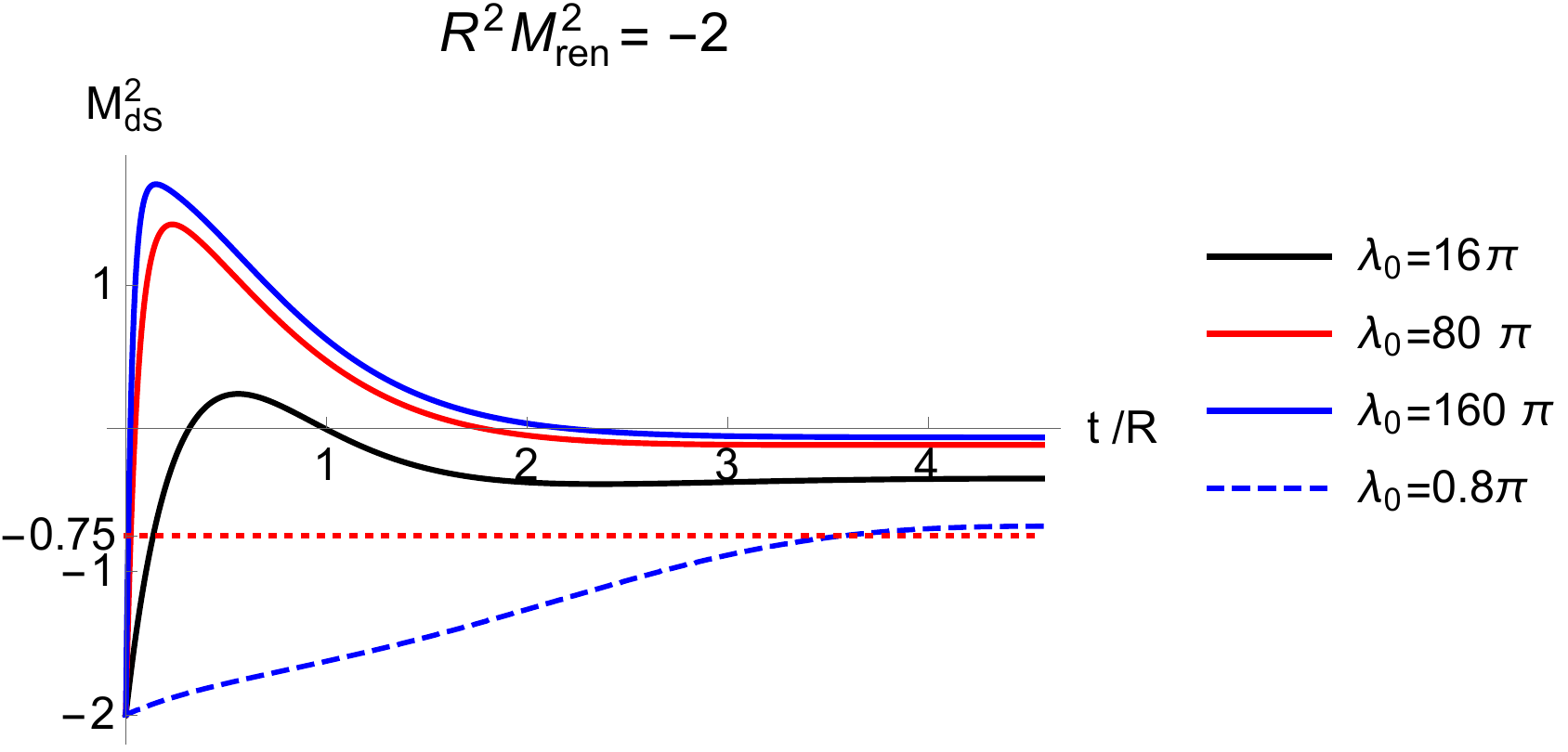}
\caption{\small{The effective mass $M_{\rm dS}$ in dS-frame as a function of de Sitter time, for different values of the interaction strength and renormalized mass $M^2_{\rm ren}=-2/R^2$.}}
\label{dsmassneg}
\end{figure}
The numerical solutions agree with this expectation: the quantum fluctuations first grow and then relax to an equilibrum value at late times. For generic values of $\lambda_0$ the system approaches a stable, free theory with (see figure \ref{dsmassneg}):
\be
\lim_{t\to \infty }M_{\rm dS}^2(t)\,=\,M_\infty^2 \,>\, -\frac{3}{4R^2}\,. 
\ee
We further find that as $\lambda_0$ is increased,  $M^2_\infty$ approaches zero.

Based on our analysis we infer that while interactions act systematically to {\em reduce} the magnitude of the effective mass in de Sitter space, it is only zero in the strict $\lambda_0 \to\infty$ limit. Therefore, except in this limit when the theory is in the vicinity of the large-$N$ Wilson-Fisher fixed point, the dynamics when transformed to E-frame is necessarily singular for generic values of the mass and quartic coupling.

It is nevertheless interesting that in the strong coupling limit the theory appears to approach a massless limit at late times, which we expect to be regular by virtue of its conformality.

\subsubsection{Initial conditions and approach to Bunch-Davies}
In this section we clarify certain  aspects of the initial conditions we have used for our system, and how they relate to standard Bunch-Davies or de Sitter-invariant boundary conditions.

The initial conditions we discussed above break de Sitter invariance, since the theory is prepared in a Gaussian, equilibrium state and de Sitter time evolution is switched on at $t=0$. 
The choice of initial states is clearly not unique, and  it is natural to consider alternate possibilities.  We would also like to understand what happens to the system if the initial state is chosen so the modes correspond to the Bunch-Davies vacuum and compare the resulting evolution with equilibrium initial conditions above. The Bunch-Davies or Euclidean vacuum is the one in which the modes are regular at one of the poles of Euclidean dS$_{3}$ ( $\simeq$ the three-sphere S$^3$). 
This has some other nice properties, to be summarised below. 
\paragraph{Bunch-Davies modes:}  
The Bunch-Davies (BD) modes for free, conformally coupled scalar fields of mass $M$, are given by
\begin{equation}
V_\ell(t) = {\rm sech}\left(\tfrac{t}{R}\right)\,\sqrt{\frac{\pi \,\Gamma{(\ell+1+\mu)}}{4 \,\Gamma{(\ell +1-\mu)}}}\, e^{i\pi\mu/2} \left(\mathrm{P}_{\ell}^{-\mu}(\tanh\tfrac{t}{R})  - \frac{2i}{\pi}\mathrm{Q}_{\ell}^{-\mu}(\tanh\tfrac{t}{R}) \right)
\end{equation} 
where $\mu = \sqrt{\tfrac{1}{4}\,-\,M^2R^2}$ and $(P_\ell^\mu, Q_\ell^\mu)$ are the associated Legendre functions.  This is the unique linear combination regular at $t/R = -i\pi/2$. The corresponding E-frame modes are 
\begin{equation}
U_\ell(\tau)\, =\, \sqrt{\cos{\tau}}\,\sqrt{\frac{\pi \,\Gamma{[\ell+1+\mu]}}{4 \,\Gamma{[\ell+1-\mu]}}}\, e^{i\pi\mu/2} \left(\mathrm{P}_{\ell}^{-\mu}(\sin{\tau})  - \frac{2i}{\pi}\mathrm{Q}_{\ell}^{-\mu}(\sin{\tau}) \right)\,.
\end{equation} 
Here, the modes are normalised so that $U_\ell\dot{U}_\ell^* - U_\ell^* \dot{U}_\ell = i$.   In the massless conformal theory when $\mu = \pm 1/2$, the modes reduce to pure exponentials:
\begin{equation}
U_\ell(\tau)= \frac{1}{\sqrt{2\ell+1}} \exp{\left(i\left(\ell+\tfrac{1}{2}\right)\left(\tfrac{\pi}{2}-\tau\right)- \tfrac{i\pi}{4}\right)}\,.
\end{equation} 
The Bunch-Davies modes at $t=0$ satisfy  the initial conditions:
\begin{align}
V_\ell(0) &= \frac{(-1)^\ell i^{-\ell} 2^{-1-\mu} \pi\,  \sqrt{\frac{\Gamma{[1+\ell+\mu]}}{\Gamma{(1+\ell-\mu)}}}}{\cos{\frac{\pi}{2}\left(\ell-\mu\right)}\,\Gamma{[\frac{1}{2}\left(1-\ell+\mu\right)]}\,\Gamma{[\frac{1}{2}\left(2+\ell+\mu\right)]}} \\
\dot{V}_\ell(0) &= \frac{i^{1+\ell} 2^{-\mu} \pi\, \sqrt{\frac{\Gamma{[1+\ell+\mu]}}{\Gamma{[1+\ell-\mu]}}}}{\sin{\frac{\pi}{2}\left(\ell-\mu\right)}\,\Gamma{\left[\tfrac{1}{2}\left(-\ell+\mu\right)\right]}\,\Gamma{\left[\tfrac{1}{2}\left(1+\ell+\mu\right)\right]}}\,.
\end{align}
When $M^2>0$, at early times, a power series expansion (in time) reveals that $|U_\ell|^2 = |U_\ell(0)|^2+h_\ell\, \tau^2$ with $h_\ell<0$ for all $\ell$. Thus the renormalized mean field, $\langle \sigma_E\rangle_{\rm ren}$ begins decreasing from its initial value at $\tau=0$. The late time value of the mean field can be calculated with Bunch-Davies initial conditions and we find (for free fields), in the small mass limit,
\be
\langle\sigma_{\rm dS}\rangle_{\rm ren}(t)\,\to\,-\frac{1}{8}M^2R\,,\qquad \qquad
\langle\sigma_{E}\rangle_{\rm ren}(\tau)\,\to\,-\frac{M^2R}{8\cos\tau}\,.
\ee
We note that our renormalized mean field is not defined with a de Sitter invariant regulator. This is the likely reason that it is actually time dependent and matches on to the de Sitter invariant value only at late times. 
\begin{figure}[h]
\centering
\includegraphics[width=3.0in]{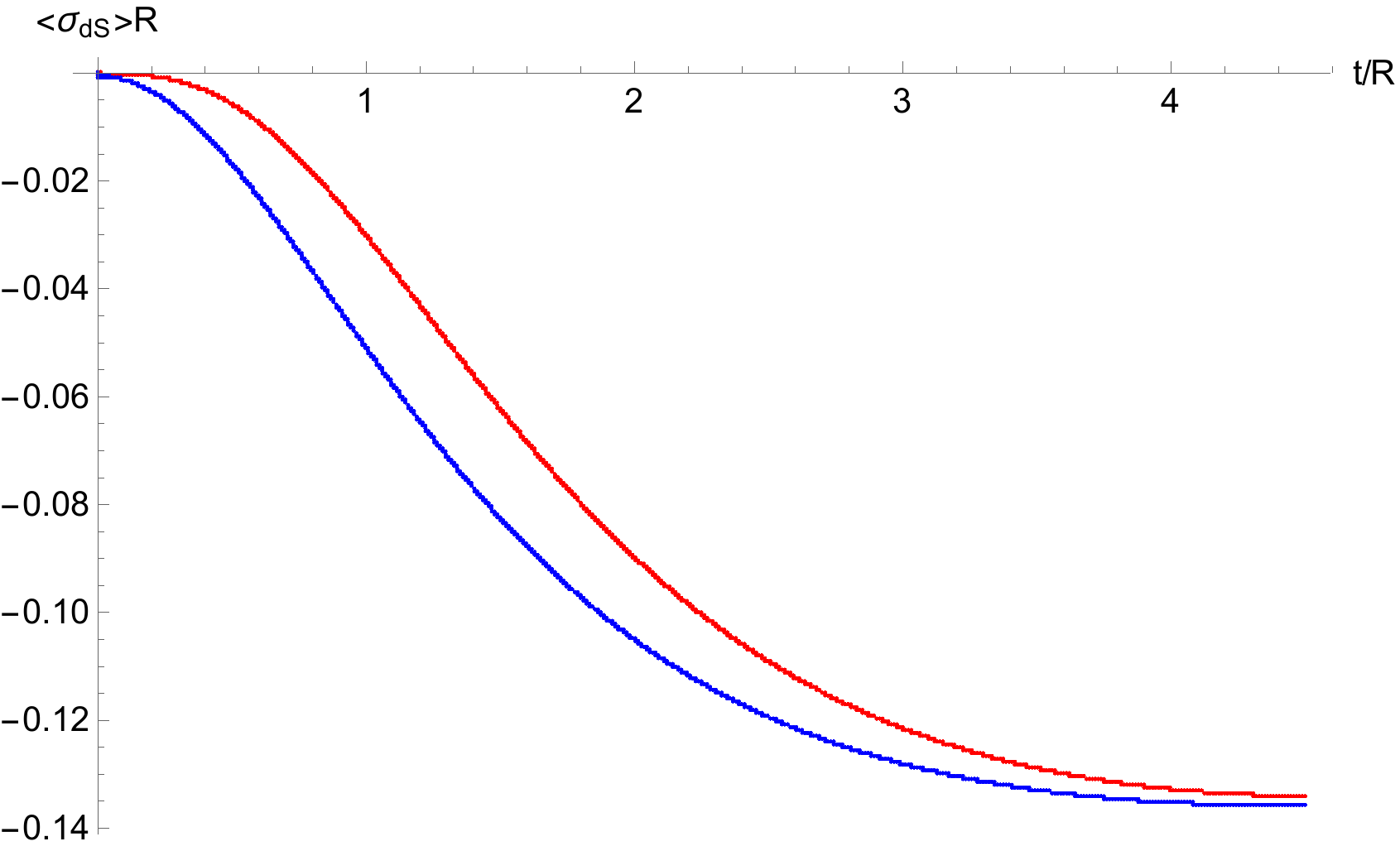}
\caption{\small{The evolution of $\langle\sigma_{\rm dS}\rangle_{\rm ren}$ for the free theory with Bunch-Davies (blue) and equilibrium (red) initial conditions, and $M^2R^2=0.1$.}}
\label{BDvseqbm}
\end{figure}

\paragraph{Equilibrium initial conditions:} For the free theory  with equilibrium initial conditions employed in this paper, the solution to the mode equations is, 
\bea
&&V_\ell(t) \,=\, {\rm sech}{\left(\tfrac{t}{R}\right)} \left(A_\ell\, \mathrm{P}_{\ell}^{-\mu}(\tanh \tfrac{t}{R}) +B_\ell\,\mathrm{Q}_{\ell}^{-\mu}(\tanh\tfrac{t}{R}) \right)\\\nonumber\\\nonumber
&&A_\ell \,=\, \frac{i (-1)^\ell 2^{\ell-\frac{3}{2}}}{\pi  \sqrt{\omega_\ell }} \Gamma{(y)} \sin{\pi\mu} \left[\omega_\ell \sin{\tfrac{\pi y}{2}}\, \Gamma{\left(\tfrac{1+x}{2}\right)} \Gamma{\left(\tfrac{1-y}{2}\right)}\,\right.\\\nonumber
&&\left.\hspace{3.5in}-\,2 i \cos{\tfrac{1}{2}\pi y}\, \Gamma{\left(1+\tfrac{x}{2}\right)} \Gamma{\left(1-\tfrac{y}{2}\right)}\right]\\
&&B_\ell= \frac{(-1)^\ell 2^{\mu-\frac{3}{2}}}{\pi^{\frac{3}{2}}\sqrt{\omega_\ell}}\sin{\pi\mu} \left[x\, \Gamma{\left(\tfrac{1+y}{2}\right)}\Gamma{\left(\tfrac{x}{2}\right)}\, -\, i\,\omega_\ell\, \Gamma{\left(\tfrac{y}{2}\right)}\Gamma{\left(\tfrac{1+x}{2}\right)}\right]\,,
\eea
where $\mu$ is defined above and 
\begin{equation}
x \,= \,\mu+\ell,\qquad y\,=\,\mu-\ell\,. 
\end{equation}
At late times we numerically observe (figure \ref{BDvseqbm}) that $\langle{\sigma}_{\rm dS}\rangle_{\rm ren}$ approaches a constant $\sigma_\infty$.  In the free field case, for small $M$ we  find,
\begin{equation*}
\sigma_\infty \,=\, - {\cal C}\, M^2R  + \mathcal{O}(M^4)\,,
\end{equation*} 
where ${\cal C}>0$ is a non-zero constant matching the Bunch-Davies value of $1/8$. At early times, 
expanding  the E-frame modes $U_\ell$ in a power series, we find that $|U_\ell|^2 \,=\, |U_\ell(0)|^2+h_\ell\, \tau^4$ with $h_\ell<0$ for all $\ell$.  Therefore $\langle{\sigma}_{E}\rangle_{\rm ren}$  decreases from its initial value, but slower than the Bunch-Davies modes.

In figure \ref{BDvseqbm} we have numerically plotted the renormalized mean field in 
the free theory for both Bunch-Davies and equilibrium initial conditions. At late times we observe that the two cases asymptote to the same value. 
\section{Discussion and future work}
This work was motivated by the existence of gravity duals of de Sitter space QFTs at large-$N$, obtained by relevant deformations of (known) CFTs yielding bulk crunch singularities in the holographic duals (e.g. \cite{hh, maldacena2, paper1, paper2}). Our goal was to understand whether the E-frame evolution of such QFTs is always singular. We picked  the $O(N)$ model in three dimensions as the simplest, nontrivial tractable example where this question could be answered by solving the theory. We have learnt that at least one (relevant) deformation of the free theory namely the quartic coupling, can result in a smooth quantum evolution, even though the microscopic coupling diverges at the end of time in the E-frame description. This regular behaviour appears to be tied to the appearance of a different  critical point  (with the large-$N$ Wilson-Fisher scaling) at the end of time. 

This suggests a potential mechanism by which big crunch bulk singularities could be avoided in special situations where the deformation drives the 
QFT towards a new fixed point as a function of time. It would therefore be interesting to find other examples where this can happen. An important point here is that any such example would require ingredients in the gravity dual that go beyond the classical (super)gravity approximation since the arguments of \cite{Abbott:1985kr} rule out any possibility of smooth evolution within ordinary gravity. It is likely that such an example will require inclusion of stringy corrections as is already suggested by the $O(N)$ model which posesses higher spin conserved currents at the critical points.

The $O(N)$ vector model also posesses marginal, sextic deformations at the free fixed point. It would be interesting to understand the effect of these in conjunction with the quartic interactions in the free theory.  It is also worth noting that although the mass deformation of the theory at finite interaction strength appeared not to posess a regular E-frame evolution, the disappearance of the mass gap at strong coupling deserves further attention. Another aspect of our analysis is that the theory is always defined in a finite volume and symmetry breaking never occurs even when the classical potential has an unstable direction (negative mass-squared) at the origin. In such cases, at late times the quantum fluctuations grow and drive the effective mass squared to positive or stable values. It would be interesting to understand the relation of this type of behaviour to the same situation considered in the  (noncompact) inflationary patch \cite{Boyanovsky:1996rw, Boyanovsky:1997xt} of de Sitter spacetime.

\acknowledgments{We would like to thank Justin David, Soo-Jong Rey and Tim Hollowood for discussions. SPK was supported by STFC grant ST/L000369/1. VV would like to thank the International Centre for Theoretical Sciences (ICTS -TIFR), Bangalore for support and hospitality, and for providing a stimulating atmosphere during the summer of 2016, when this work was commenced.

\end{document}